\documentclass{agujournal}
\usepackage{url}
\usepackage{color}
\usepackage{combelow}
\usepackage[normalem]{ulem}

\journalname{JGR-Space Physics}

\begin{document}

%
%


\title{Properties and geoeffectiveness of solar wind high-speed streams and stream interaction regions during solar cycles 23 and 24}

%
%




\authors{Maxime Grandin\affil{1,2}, Anita~T. Aikio\affil{3}, and Alexander Kozlovsky\affil{1}}


\affiliation{1}{Sodankyl\"a Geophysical Observatory, University of Oulu, Sodankyl\"a, Finland}
\affiliation{2}{Department of Physics, University of Helsinki, Helsinki, Finland}
\affiliation{3}{Ionospheric Physics Unit, University of Oulu, Oulu, Finland}





\correspondingauthor{Maxime Grandin}{maxime.grandin@helsinki.fi}




\begin{keypoints}
\item Stream interaction regions and high-speed streams (SIR/HSSs) are 20--40\% less geoeffective during solar cycle (SC) 24 than during SC23
\item The most geoeffective SIR/HSSs in solar cycles 23 and 24 take place in the early declining phases
\item During the late declining phase of SC23, both SIR/HSS event number and maximum velocity are highest, yet their geoeffectiveness is low
\end{keypoints}

%
%


\begin{abstract}
We study the properties and geoeffectiveness of solar wind high-speed streams (HSSs) emanating from coronal holes and associated with stream interaction regions (SIRs). This paper presents a statistical study of 588~SIR/HSS events with solar wind speed at 1~AU exceeding 500~km/s during 1995--2017, encompassing the decline of solar cycle~22 to the decline of cycle~24. Events are detected using measurements of the solar wind speed and the interplanetary magnetic field (IMF). Events misidentified as or interacting with interplanetary coronal mass ejections (ICMEs) are removed by comparison with an existing ICME list. Using this SIR/HSS event catalog (list given in the supplementary material), a superposed epoch analysis of key solar wind parameters is carried out. It is found that the number of SIR/HSSs peaks during the late declining phase of solar cycle (SC)~23, as does their velocity, but that their geoeffectiveness in terms of the $AE$ and $SYM$--$H$ indices is low. This can be explained by the anomalously low values of magnetic field during the extended solar minimum. Within SC23 and SC24, the highest geoeffectiveness of SIR/HSSs takes place during the early declining phases. Geoeffectiveness of SIR/HSSs continues to be up to 40\% lower during SC24 than SC23, which can be explained by the solar wind properties.
\end{abstract}

\section{Introduction}

Corotating solar wind high-speed streams (HSSs) were discovered by \citet{Snyder1963}, who showed a correlation between recurring geomagnetic activity and the solar wind speed measured by Mariner~2. High-speed streams were later associated with solar coronal holes \citep[e.g.,][]{Krieger1973,Sheeley1976}. Since a given coronal hole may persist for several solar rotations, this will lead to recurring geomagnetic activity with a $\sim$27-day period in many parameters in the atmosphere, ionosphere, and magnetosphere \citep[e.g.,][]{Chkhetiia1975,Hapgood1993,Temmer2007,Crowley2008}, this periodicity corresponding to the synodic solar rotation period as viewed from Earth. A HSS is characterized by an enhancement in the solar wind velocity lasting for several days. Typically, the speed exceeds 500~km/s for two to three days, and may reach a maximum above 800~km/s \citep[e.g.,][]{Denton2012,Kavanagh2012}. 

When the high-speed stream overtakes the slow-speed background stream, this leads to compressions of both the interplanetary magnetic field (IMF) and the plasma density in the rising-speed portion of the high-speed stream \citep{Belcher1971}. The resulting structure was called corotating interaction region (CIR) \citep{Smith1976,Gosling1999}. In this paper, we refer to this interaction region as ``stream interaction region (SIR)'' regardless of the number of solar rotations during which it is observed. This practice is similar to, e.g., \citet{Richardson2018}. In addition, the high-speed stream events that we study contain the interaction regions in the leading edge of the streams, so we refer to these events as SIR/HSS.

SIR/HSSs have been particularly studied during the declining phase of the 11-year solar cycle, since at that time they represent the main cause for geomagnetic disturbances \citep{Gonzalez1999,Tsurutani2006}. On the other hand, during the maximum years of solar activity, coronal mass ejections (CMEs) are the major cause of geomagnetic disturbances \citep[e.g.,][]{Richardson2012_storms}. CMEs are eruptions of magnetized plasma from the solar atmosphere, with a broad range of propagation speeds \citep{Gopalswamy1992}. Interplanetary coronal mass ejections (ICMEs) are the interplanetary counterpart of CMEs \citep[e.g.,][]{Schwenn1983,Sheeley1985,Lindsay1999,Webb2000}. Those produce the most extreme space weather events \citep[e.g.,][]{Richardson2000,Richardson2001,Zhang2007,Echer2013}.

It is well-established that SIR/HSSs are predominantly responsible for weak to moderate geomagnetic storms \citep[e.g.,][]{Tsurutani2006}. \citet{Zhang2008} found that about 80\% of the geomagnetic storms produced by ``pure'' SIR/HSSs (i.e., SIR/HSSs which are not interacting with an ICME) are weak ($-50 < \mathrm{Dst} \leq -30$~nT) to moderate ($-100 < \mathrm{Dst} \leq -50$~nT), while \citet{Alves2006} showed that only about 33\% of the SIR/HSSs are responsible for moderate to intense storms ($\mathrm{Dst} \leq -100$~nT). \citet{Chi2018} found that a large fraction of the most intense SIR/HSS-related geomagnetic storms are produced by SIR/HSSs interacting with an ICME. According to \citet{Kilpua2017}, although SIR/HSSs are frequent during solar minimum, they produce fewer geomagnetic storms as they are generally characterized with lower speed gradients, lower dynamical pressure peaks and lower interplanetary magnetic field magnitude. Yet, \citet{Richardson2012_storms} revealed that almost half of the large storms during solar minimum are due to SIR/HSSs.

Statistical studies of SIR/HSS effects on the magnetosphere--ionosphere--thermosphere system have shown dropouts of relativistic electrons in the outer radiation belt \citep{Borovsky2009,Morley2010}, followed by a recovery to higher flux levels in the case of strong events \citep{Denton2012}, enhancement of substorm activity \citep{Tsurutani2006}, depletion of the ionospheric $F$ region \citep{Denton2009,Grandin2015}, energetic ($E>30$~keV) electron precipitation into the $D$ region \citep{Meredith2011,Grandin2017}, and enhancement of ULF wave activity in the magnetosphere \citep{Mathie2001}.

Because of the strong influence of SIR/HSS events on the planetary environment, objective criteria to identify SIR/HSS events from solar wind data would be very useful. Previous statistical studies have used methods to detect SIR/HSSs focusing on stream interfaces, and hence searching for not only enhancement in the solar wind radial velocity but also a west--east deflection of the plasma flow \citep[e.g.,][]{Morley2010,Denton2012,Kavanagh2012}. Those methods often required a visual inspection of the events and/or comparison with existing lists of other solar wind disturbances such as ICMEs to remove false positives. For instance, \citet{Jian2006} produced a fairly comprehensive list of events during 1995--2004, which was extended till the end of 2016 by \citet{Chi2018}. The events were picked by eye based on criteria on the solar wind speed and total perpendicular pressure. The starting time corresponds to the time when the total perpendicular pressure starts increasing, which is close to the time when the speed also starts increasing. \citet{Richardson2012_flows} have also produced a classification of solar wind flows, including an identification of SIR/HSSs, during 1963--2011 based on visual inspection of solar wind data as well as geomagnetic observations, energetic particle data, and neutron monitor data. Others have developed automated methods using only criteria related to the enhancement of the solar wind speed, such as Mari\cb{s} Muntean et al. (\url{http://www.geodin.ro/varsiti/}) during 2009--2016 or \citet{Gupta2010} during 1996--2007.

More generally, attempts to automatically identify solar wind flow types between interstream flow (slow wind), coronal-hole flow (high-speed streams), and transient (ICMEs) have been presented by, e.g., \citet{Zhao2009} during 1998--2008 from ACE data, \citet{Reisenfeld2013} during the Genesis mission (2001--2004), and \citet{Xu2015} during 1963--2013 from OMNI data. However, when comparing results from the latter three identification schemes, \citet{Neugebauer2016} found that the overall agreement is limited, with all three algorithms giving a same flow type only 49\% of the time.

This paper presents a method to identify SIR-associated HSSs, which is applied to 23 years of solar wind observations from 1995 until 2017. This detection method is based on the one used to obtain the HSS events analyzed in \citet{Grandin2015,Grandin2017} for 2006--2008, and for this paper some improvements have been developed, detailed in section~3. The aim of this method is to find the time when the SIR region with compressed magnetic field hits the bow shock. Solar wind speed reaches the maximum typically 1--2 days after this zero epoch. It was shown in \cite{Grandin2015,Grandin2017} that the zero epoch selected in this manner corresponds very well to the time when geomagnetic activity in terms of the AE index starts to increase. The list of SIR/HSSs obtained with this new algorithm is used to self-consistently study the features of SIR/HSSs impinging on the Earth's magnetopause during the different phases of two solar cycles, and in specific their geoeffectiveness.

This paper is organized as follows: the data sets used in this study are presented in section~2, and the algorithm developed to gather the list of SIR/HSS events is described in section~3. Section~4 presents the results on SIR/HSS characteristics from the end of solar cycle 22 (SC22) until late 2017 (declining phase of SC24). Section~5 discusses the results and compares the obtained SIR/HSS list with other existing ones, and section~6 summarizes the main conclusions.

\section{Data}
The data used in this study consist of interplanetary magnetic field (IMF) and solar wind data measured near the Earth between 1995 and 2017. The data were obtained by several satellites of which the main ones are WIND (1995--present) \citep{Ogilvie1995} and ACE (1998--present) \citep{McComas1998,Smith1998}. The measured data are available through the OMNI database \citep{King2005}, where they have already been propagated to the terrestrial bow shock. The parameters of interest in this study are the IMF magnitude $B$, the solar wind speed $V$, the solar wind density $N$, and the Akasofu $\varepsilon$ parameter \citep{Akasofu1979} calculated using
\begin{equation}\label{eq:Akasofu}
\varepsilon = \dfrac{4\pi}{\mu_0} V B^2 L_0^2 \sin^4 \left( \dfrac{\theta}{2} \right),
\end{equation}
with $\mu_0$ the vacuum permeability, $\theta$ the IMF clock angle in the plane perpendicular to the Sun--Earth direction, and $L_0 \simeq 7 R_E$. The Akasofu $\varepsilon$ parameter is a coupling function often used as a proxy for energy input from the solar wind into the magnetosphere. For this study, the OMNI data were retrieved for years 1995--2017 at 1~h time resolution.

We also take the 1--h averages of the auroral electrojet ($AE$) and \emph{SYM--H} geomagnetic indices \citep{Davis1966,Iyemori1990} from OMNI to monitor the response of the geospace in terms of substorm and storm activity, respectively.

In addition, in order to discuss the properties of SIR/HSSs during different phases of solar cycles 23 and 24, the monthly values of the revised sunspot number between 1995 and 2017 have been gathered from the World Data Center Sunspot Index and Long-Term Solar Observations (SILSO) \citep{sidc}.

\section{SIR/HSS Event Detection Algorithm}

Our aim is to find SIR/HSS events that can be geoeffective. Therefore, for maximum velocity we have chosen a threshold of 500~km/s. One can see that the Akasofu epsilon solar wind coupling function (eq.~1) depends linearly on solar wind velocity. The papers by \citet{Kavanagh2012} and \citet{Denton2012} are also in accordance with this selection, see point 2(c) below. However, the algorithm presented below could be adapted to also detect SIR/HSSs of lower maximum speed.

The algorithm used to detect SIR/HSS events for this study consists of four steps. The first three steps each consider a criterion based on solar wind data at 1~h resolution provided by OMNI.
\begin{enumerate}
\item The SIR is formed by the plasma compression arising from the interaction of the slow and fast solar wind streams. Therefore the leading edge of SIRs exhibits an enhancement in the IMF magnitude $B$ \citep[e.g.,][]{Richter1986,McPherron2006,Denton2012}. Hence, as a first criterion, the time derivative of $B$ is estimated at each time $t_i$ using
\begin{equation}
\dfrac{\mathrm{d}B}{\mathrm{d}t} \, \simeq \, \dfrac{1}{2 \Delta t} \left( B_{i+1} - B_{i-1} \right),
\end{equation}
with $\Delta t = t_{i+1} - t_i=1$~h the time resolution of the solar wind data. Each instant $t_i$ at which this time derivative value exceeds the empirically determined threshold of 0.6~nT/h is then flagged as a candidate SIR/HSS event starting time ($t_0$). If the threshold value is too large, only SIRs that are associated with a forward shock will be detected. \citet{Jian2006} found that, on average, the occurrence rate of forward shocks in SIRs at 1 AU is about 18\%. However, if the threshold value is too small, then we are no more looking at compressed plasma, but some natural fluctuations in B. The threshold value used in this paper was found empirically, and comparison to other existing SIR lists described in Section 5.1 shows that the threshold works well, since no events are missed due to this criterion.

\item The second criterion is based on the solar wind velocity $V$. The candidate events obtained with criterion (1) are tested for three characteristics:
  \begin{enumerate}
  \item To avoid detecting events taking place while the background solar wind speed is already high, at time $t_0$, one must have $V(t_0) \le 450$~km/s, bearing in mind that occasional compound events consisting of two high-speed streams interacting with each other might not be detected due to this restriction. This value of 450~km/s was also determined empirically by visual inspection of the detected/undetected SIR/HSS events when trying various velocity thresholds.
  \item To ensure that the velocity starts increasing shortly after the beginning of the event, the average velocity slope between $t_0$ and $t_0+1$~day must be greater than 30~(km/s)/day, which is consistent with definitions of HSSs adopted by \citet{Bame1976} and \citet{Gosling1976}.
  \item To keep only events with solar wind speed significantly higher than the slow wind, a threshold is defined: the solar wind speed must reach at least 500~km/s within [$t_0$,$t_0+3$~days]. This value of 500~km/s is the same as used by \citet{Kavanagh2012} in their superposed epoch study of $>$30~keV precipitation during high-speed streams, and falls between the thresholds defined by \citet{Denton2012} for ``weak'' (450~km/s) and ``strong'' (550~km/s) HSS events. It is also consistent with the retained velocity thresholds used for the detection of coronal-hole flow by the Genesis spacecraft mission \citep{Reisenfeld2013}.
  \end{enumerate} 
The candidate SIR/HSS event starting times which verify each of those three conditions are kept for the next step.

\item The third criterion makes sure that one does not detect a same SIR/HSS event twice when, for instance, criteria (1) and (2) are met at several times during the event. Criterion (3) consists in keeping only SIR/HSS event starting times separated by at least 3~days. Beginning with the first detected SIR/HSS event at $t_0$, all candidate times between $t_0$ and $t_0+3$~days are examined, and the one associated with the lowest solar wind speed, at time $t_0^\prime$, is retained while all others are removed from the list. This guarantees that the solar wind speed starts increasing immediately after the retained starting time. This operation is repeated for the next SIR/HSS candidate occurring after $t_0^\prime+3$~days, and so on until the end of the list of candidates is reached.
\end{enumerate}

When applied to the solar wind data collected between 1~January~1995 and 31~December~2017, these three criteria lead to the detection of 709~candidate SIR/HSS events. However, some of these events are not purely SIR-related HSSs.
\begin{enumerate}
\setcounter{enumi}{3}
\item The list of candidates is then compared to the list of near-Earth ICMEs provided by \citet{Richardson2010}, which covers years from 1996 onward and is regularly updated (\url{http://www.srl.caltech.edu/ACE/ASC/DATA/level3/icmetable2.htm}). A candidate is considered a likely ICME or contaminated by an ICME if one event from the Richardson and Cane list with $V_\text{max} \geq 500$~km/s reached the Earth between $t_0-1$~day and $t_0+3$~days, $t_0$ being the date of the candidate SIR/HSS event. 
\end{enumerate}

In total, 122~events matching criterion~4 were found and were removed from the SIR/HSS catalogue. As a consequence, this leaves a total of 588~SIR/HSS events obtained with this four-step method. This list of SIR/HSS events is provided as supporting information. The list gives for each SIR/HSS event its starting time $t_0$, initial solar wind speed, the time and value of maximum solar wind speed during [$t_0$, $t_0$+3~days], and an end time defined as the first occurrence when the solar wind speed drops below 450~km/s after reaching its maximum. All times are given in UT for solar wind propagated at the terrestrial bow shock as given in OMNI. It should be noted that this list does not claim to be comprehensive, as (i) this study focuses on SIR/HSS events whose velocity reaches at least 500~km/s, (ii) occasionally, some SIR/HSS events with $V_\textbf{max} \geq 500$~km/s might have been undetected by the method, and (iii) events containing ICMEs during [$t_0-1$~day, $t_0+3$~days] are removed.

\begin{figure}
 \centering
 \includegraphics[width=\textwidth]{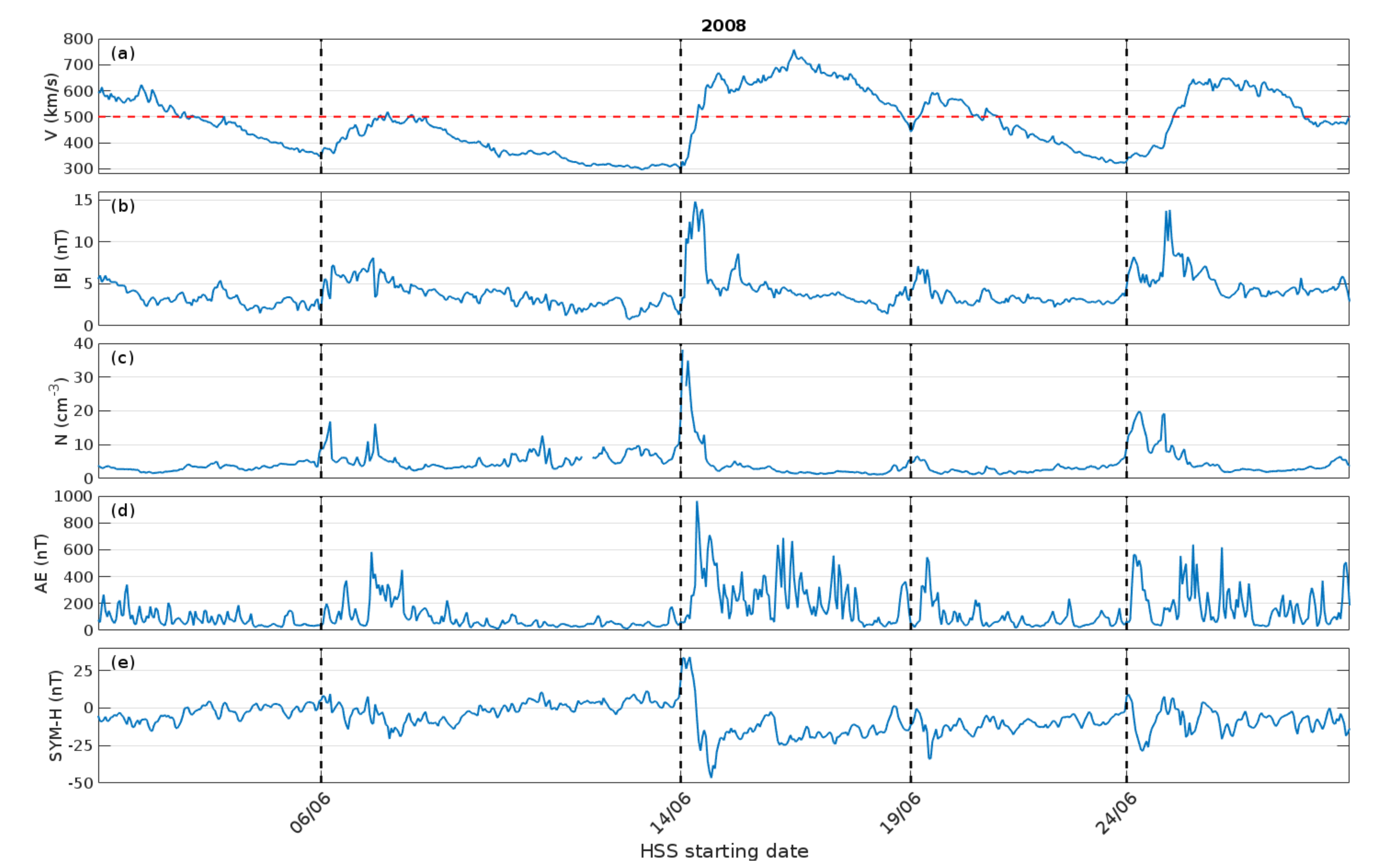}
 \caption{SIR/HSS events detected by the algorithm in June~2008 with a maximum solar wind velocity greater than 500~km/s. Each vertical dashed line corresponds to the beginning of an event. The time series shown are (a) the solar wind velocity, (b) the IMF magnitude, (c) the solar wind number density, (d) the $AE$ index, and (e) the \emph{SYM--H} index. The horizontal dashed red line in the first panel indicates $V=500$~km/s, used for criterion~2(c).}
 \label{fig1}
\end{figure}

Figure~\ref{fig1} illustrates some aspects of the detection algorithm on solar wind data gathered in June 2008. During this time window, four SIR/HSS events were detected by the algorithm. Their starting times are 6~June at 4 UT, 14~June at 12 UT, 19~June at 20 UT, and 24~June at 20 UT, and those are indicated by vertical dashed lines and the ticks on the $x$-axis. Figure~1a shows the solar wind speed. It is clear that the beginning of each event is set very close to the time when the solar wind velocity starts to increase, which naturally comes from criteria~2b and 3. The horizontal red dashed line indicates the threshold value at 500~km/s for a HSS event (criterion 2c). Figure~1b gives the IMF magnitude $|B|$ which is used for criterion~1. The sharp time derivative used for the detection is particularly prominent at the beginning of the second event. From Figure~1c, which shows the solar wind number density, it can be noted that the beginning of the detected SIR/HSS events is close to the maximum of the density peak (again, this is more prominent for the second event), which corresponds to earlier findings \citep[e.g.,][]{Gosling1972,McPherron2006,Denton2009}.
 Figure~1d displays the $AE$ index, which is a proxy for substorm activity \citep{Iyemori1990}. Substorm activity starts to increase very close to the determined onset time of the events in each case. It is interesting to note that AE starts to increase well before the solar wind velocity has reached a high value, which indicates that the SIR is geoeffective. Figure~1e displays the \emph{SYM--H} index, which describes the intensity of the storm time ring current \citep{Wanliss2006}. In the 14~June event, which is associated with the strongest storm, the zero epoch matches very well with the sudden storm commencement signature showing as a positive peak in \emph{SYM--H}, likely to be produced by the slight dayside magnetopause compression associated with the SIR arrival. A smaller positive peak is observed also for the other events, except the first one on 6 June. All these events are associated with weak magnetic storms. It is well-known that SIR-associated geomagnetic storms are frequent in the declining phase of the solar cycle and are typically weak to moderate \citep{Tsurutani2006,McPherron2006,Richardson2012_storms,Grandin2017}.

\begin{figure}
 \centering
 \includegraphics[width=\textwidth]{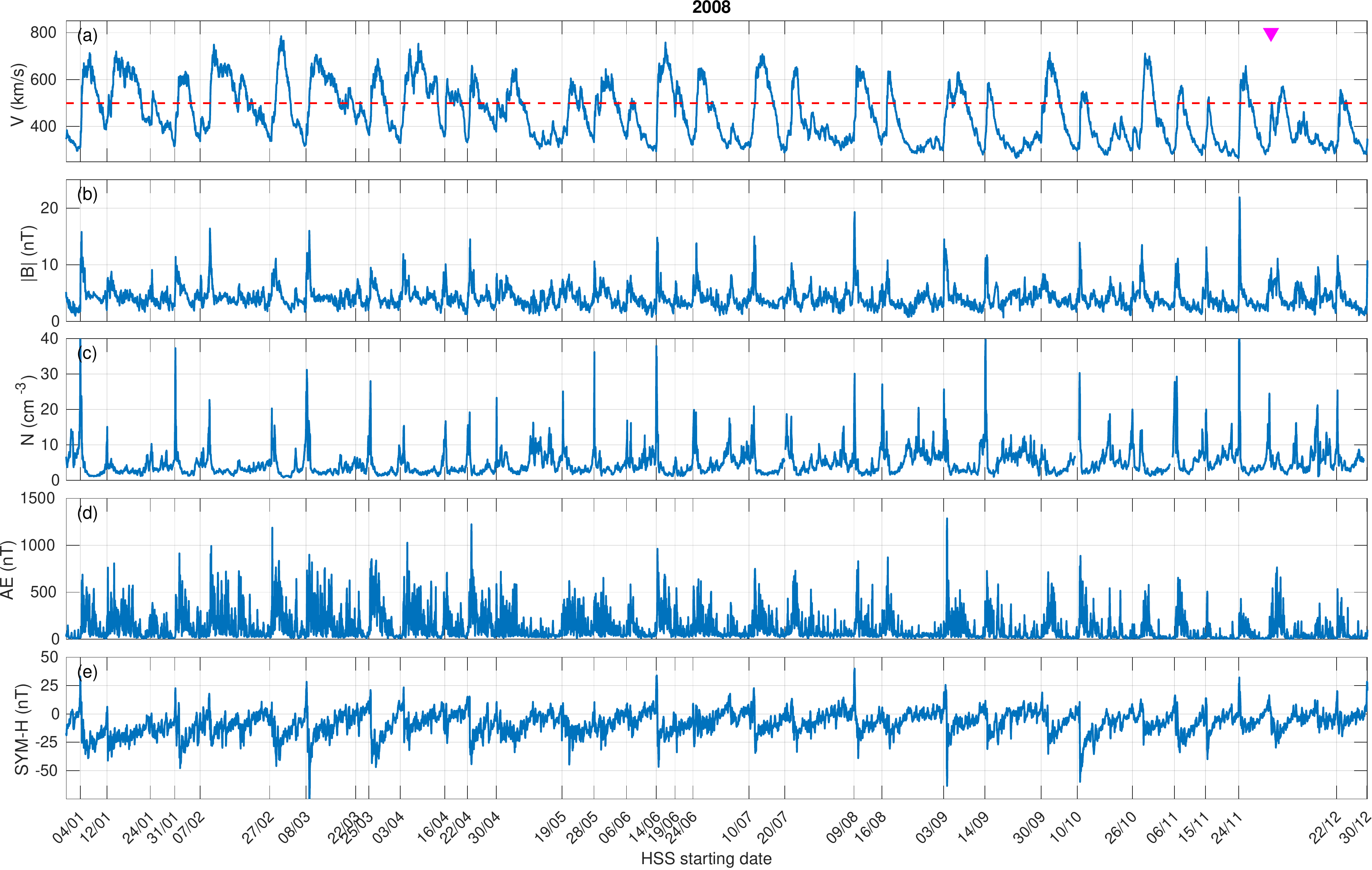}
 \caption{Same as Figure~\ref{fig1} but showing the data and detected events for the entire year 2008. The magenta triangle in panel (a) corresponds to the ICME falsely detected as SIR/HSS in 2008, which was removed when applying criterion~4 (see Figure~3).}
 \label{fig2}
\end{figure}

Figure~\ref{fig2} shows the same data as Figure~\ref{fig1} but for the whole year 2008, during which 33~SIR/HSS events were detected by the algorithm. From this longer time series, the features described above can be identified in most of the detected events, namely the fact that the beginning of events corresponds well with the start of the solar wind speed increase, with a peak in the solar wind density within the SIR region, with the start of AE index enhancement, and is near the sudden storm commencement signature when it is present. One can note that the event contaminated by the only ICME with $V_\text{max} \geq 500$~km/s in 2008, which took place on 4~December (magenta triangle in Figure~2a), was rejected by the algorithm because of criterion~4.

\begin{figure}
 \centering
 \includegraphics[width=.8\textwidth]{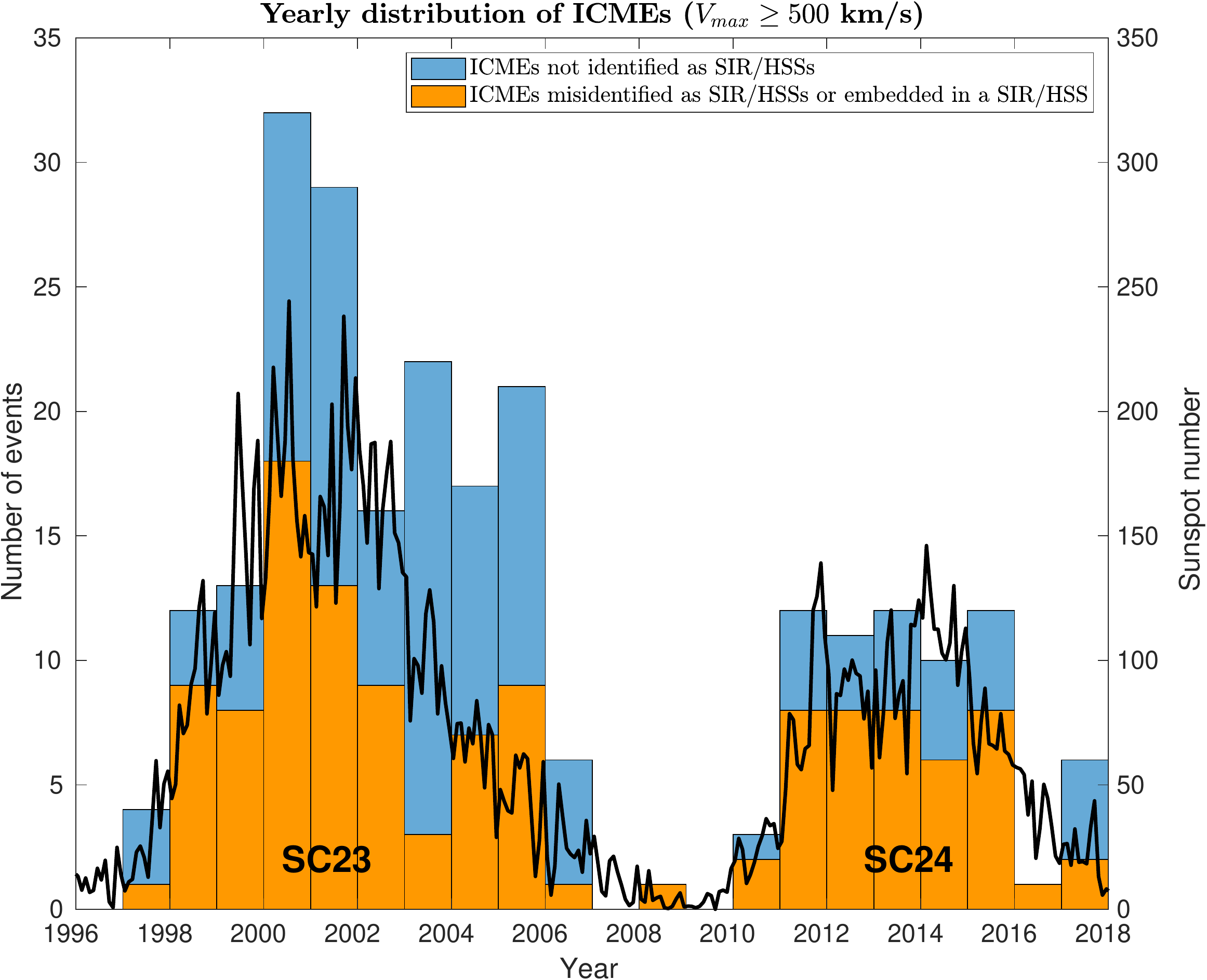}
 \caption{Histogram, left axis: Yearly distribution of ICMEs from \citet{Richardson2010} (updated online version of January 2019) with ${V_\text{max} \geq 500}$~km/s during 1996--2017 (envelope) and, among these, ICMEs misidentified as SIR/HSSs by the algorithm or ICMEs embedded in a SIR/HSS (orange). Black line, right axis: Monthly sunspot number during those years.}
 \label{fig3}
\end{figure}

Figure~\ref{fig3} gives the yearly distribution of ICMEs with {$V_\text{max} \geq 500$~km/s} from the Richardson and Cane list between 1996 and 2017 (the envelope of the histogram). Orange bars show the ICME events that were identified as SIR/HSSs by our algorithm with steps 1--3, before they were removed based on criterion 4, and blue bars show the remaining ICME events. The black line gives the monthly sunspot number during those years. It is noteworthy that not all the ICMEs are misidentified as SIR/HSSs by our algorithm, only 122 out of 237 ICMEs with speeds larger than 500~km/s, corresponding to about 50\%. Of all the detected SIR/HSS events, the amount of rejected events (ICMEs or SIR/HSSs contamined by ICMEs) is 17\% (122/709). This underlines the fact that comparison with an ICME list (criterion~4) is a necessary step in the algorithm, as the solar wind parameters used in the identification scheme during ICMEs quite often exhibit similar features as SIR/HSSs. This issue is not severe during solar minimum years, but becomes important during solar maximum years. One must also keep in mind that some of the rejected events may be real SIR/HSS events that are interacting with a simultaneously occurring ICME event near the ecliptic plane \citep[see, e.g.,][]{AlShakarchi2018,Shugay2018}.

\section{Results}

\subsection{Overall Features of Detected Events}

\begin{figure}
 \centering
 \includegraphics[width=0.8\textwidth]{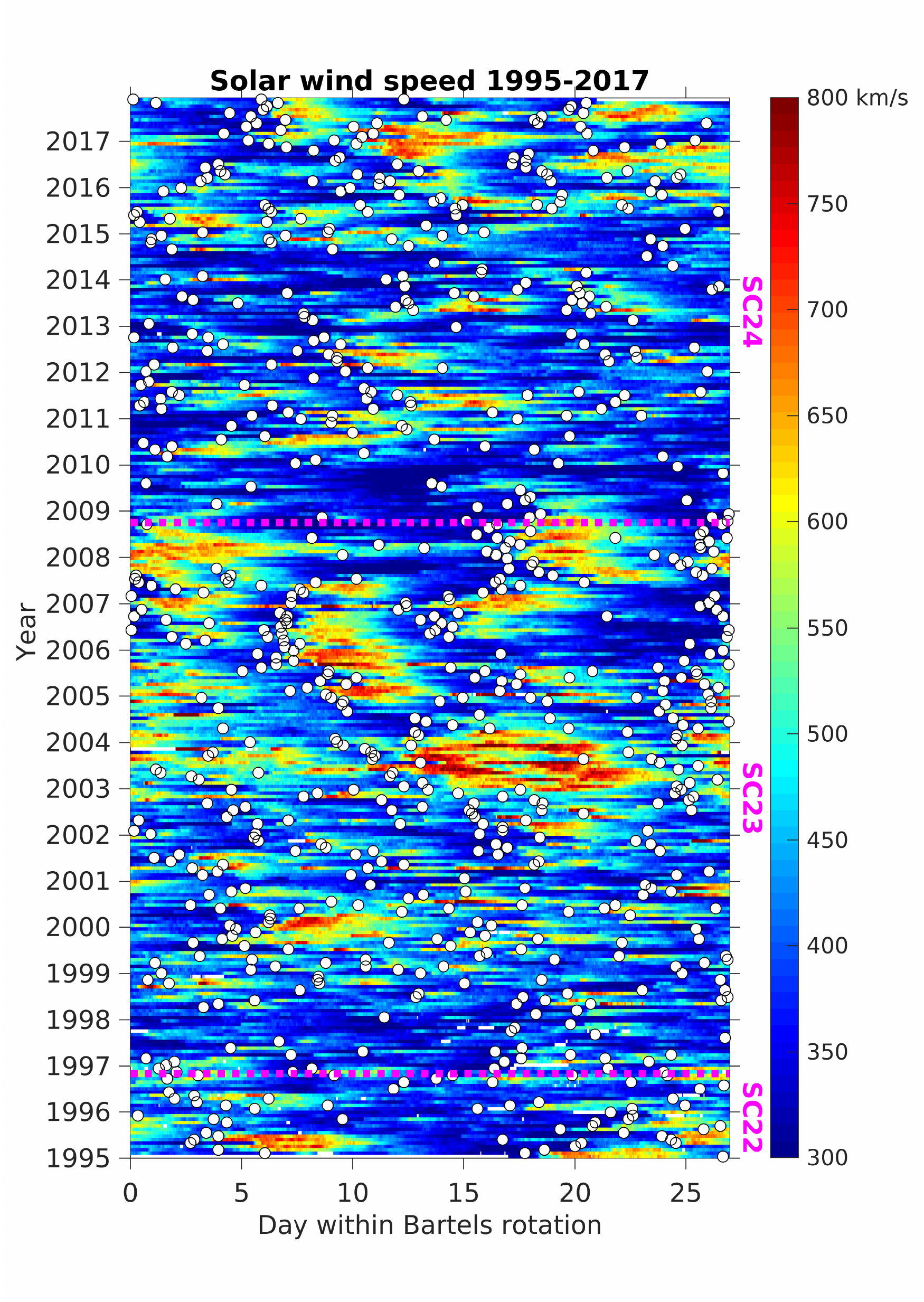}
 \caption{Color plot: Solar wind velocity plotted as a function of day number within 27-day Bartels rotation ($x$-axis, 1~Jan~1995 corresponds to day 17 of Bartels rotation 2204) and year from 1995 to 2017 ($y$-axis). White circles: starting dates of the detected SIR/HSS leading edges. The magenta dashed lines indicate the boundaries between solar cycles obtained from  \url{ftp://ftp.ngdc.noaa.gov/STP/space-weather/solar-data/solar-indices/sunspot-numbers/cycle-data/table_cycle-dates_maximum-minimum.txt}.}
 \label{fig4}
\end{figure}

In order to provide an overview of all the detected events during 1995--2017, Figure~\ref{fig4} displays the solar wind velocity plotted as a function of year ($y$-axis) and day of solar rotation ($x$-axis), when using Bartels rotations (27-day period). The time series starts from 1~January~1995 on day~17 of Bartels rotation 2204 and follows the horizontal axis until day~27, and then continues above from day~0 of Bartels rotation 2205 which starts on 11~January, and so on. The $y$-axis however indicates year instead of Bartels rotation number for an easier comparison with the other figures. This way of displaying the solar wind speed enables one to identify features recurring over several solar rotations, which appear as vertical structures in the plot. Two horizontal magenta dashed lines, in August~1996 and December~2008, indicate the transitions between solar cycles 22 and 23, and solar cycles 23 and 24, respectively, obtained from \url{ftp://ftp.ngdc.noaa.gov/STP/space-weather/solar-data/solar-indices/sunspot-numbers/cycle-data/table_cycle-dates_maximum-minimum.txt}. The starting dates of the detected SIR/HSS events are indicated as white filled circles on top of the solar wind velocity color plot. Several features indicating recurring elevated solar wind velocities can be identified, in particular during year 1995, years 2003--2004, years 2003--2008, and from 2015 onwards. During one solar rotation period, there may be several recurring intervals of high-speed flows. Discussion of the origin of these long-lived recurring high-speed streams is beyond the scope of the present paper, but one can note that, e.g., \citet{Temmer2007} suggest that the 9-day period in solar wind parameters, showing up as higher harmonic of the solar rotation frequency, is caused by the ``periodic'' longitudinal distribution of coronal holes on the Sun recurring for several solar rotations, especially during January--September 2005. This 9-day periodicity can be seen in Figure~\ref{fig4} during that time period, with enhanced solar wind speed recurring around Bartels rotation day~0, day~9, and day~18. It is noteworthy that structures from which high-speed solar wind flows may remain present at a same phase of the Bartels rotation for up to several years (see, e.g., the high-velocity signatures near the longitude corresponding to day~10, recurring from early 2005 until late 2006, or those near the longitude corresponding to day~2, recurring from mid 2006 until late 2008). This is consistent with findings by \citet{HeidrichMeisner2017} who studied a same coronal hole structure for twelve solar rotations in 2006. The resulting HSS reached the Earth eleven times, yet exhibiting variability in its signature observed by ACE at L1 as the spacecraft mapping drifted to different regions within the coronal hole. On the other hand, \citet{Krista2018} tracked equatorial coronal holes at the surface of the Sun during 2011--2014 using the unique 360$^\circ$ coverage offered by the configuration of the Solar--Terrestrial Relations Observatory (STEREO) constellation and the Solar Dynamics Observatory (SDO). During those years in the maximum phase of SC24, most equatorial coronal holes had a lifetime of less than 100~days ($<$4 solar rotations), which is consistent with the significantly smaller number of recurring HSSs observed at L1 during 2011--2014 compared to 2006, as can be seen in Figure~\ref{fig4}.

During the declining phases of these solar cycles, the detected SIR/HSS events appear to be often recurring. During the rising phase of solar cycles~23 and 24, recurring HSS signatures are less prominent, and the detected events tend to occur at more random solar longitudes. This is in agreement with findings by \citet{Borovsky2006a} for the end of SC22 and the beginning of SC23. The fading of the recurring coronal hole signatures coincides quite strikingly with the transition between solar cycles 23 and 24 in late 2008, after which HSSs become less frequent for a couple of years.

\begin{figure}
 \centering
 \includegraphics[width=.8\textwidth]{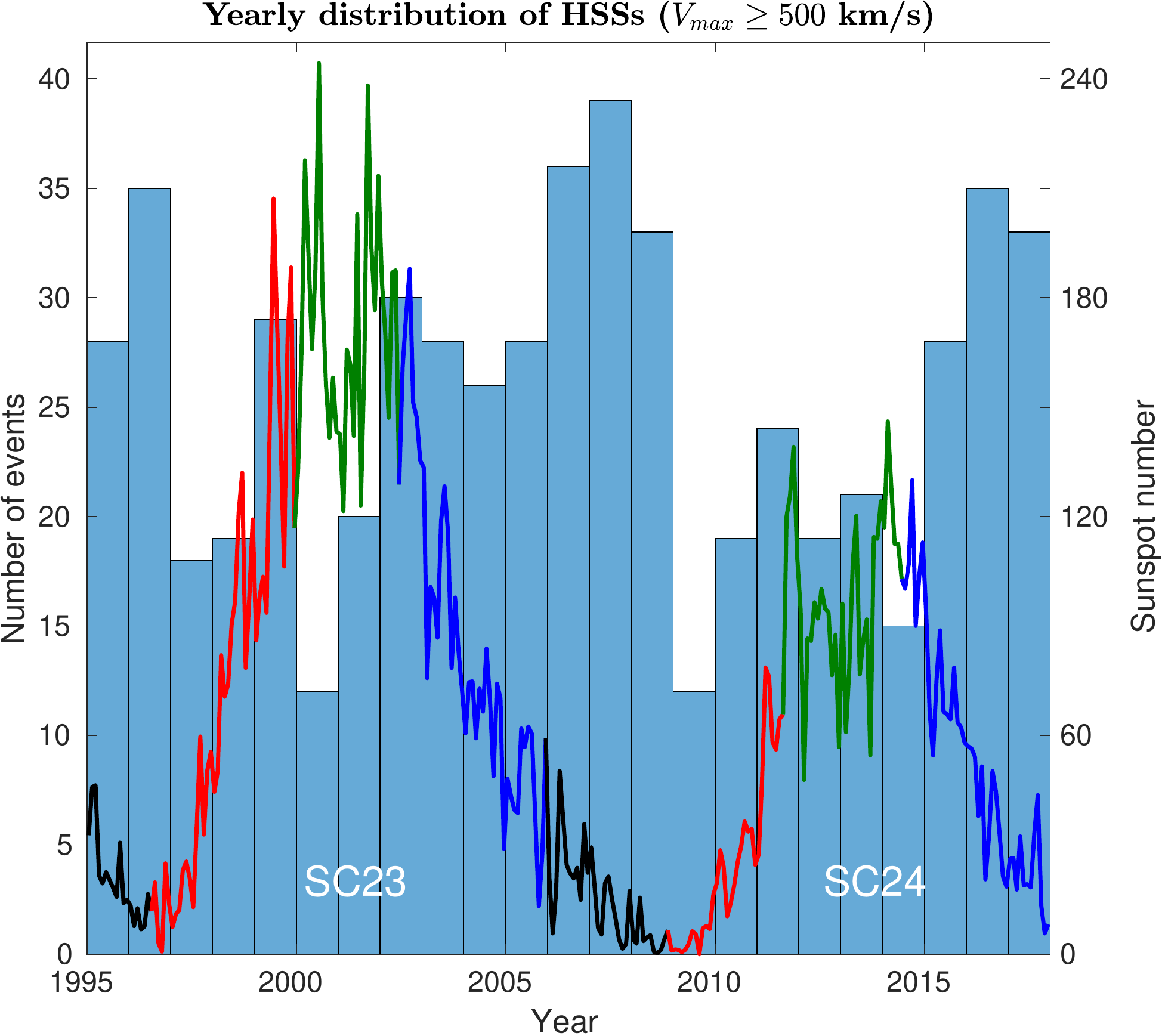}
 \caption{Histogram: Yearly distribution of high-speed stream events detected with the method using the criterion {$V_\text{max}\ge 500$~km/s}. Line plot: Monthly sunspot number. The colors of the curve indicate the time intervals which are referred to as ``rising phase'' (red), ``maximum'' (green), ``early declining phase'' (blue) and ``late declining phase'' (black) of solar cycles.}
 \label{fig5}
\end{figure}

Figure~\ref{fig5} shows the yearly number of SIR/HSS events with $V_\mathrm{max} \ge 500$~km/s detected by our method as a histogram plot, where all the events that are contaminated by ICMEs have been removed. The sunspot number is shown as a colored curve, which corresponds to four phases of solar cycles which we will consider below (in particular in section~4.2): rising phase (red), maximum (green), early declining phase (blue), and late declining phase (black). The selected time limits for the phases are given in Table~\ref{table1}. The criterion used to separate the rising, maximum and declining phases was that, for both cycles~23 and 24, the double peak of the monthly sunspot number be included in the maximum phase. The motivation for subdividing the declining phase of solar cycles into early and late parts was to obtain solar cycle divisions of roughly similar durations, and to reveal potential differences within the declining phase itself.

While it is well-known that high-speed streams are frequent during the declining phases of solar cycles \citep[e.g.,][]{Gonzalez1999}, many events are also detected during the other phases of solar cycles 23 and 24, which is consistent with reports by \citet{Richardson2012_flows} during solar cycles 20 to 23. It is noteworthy that even the rising phases contain a large number of SIR/HSS events, corresponding to 69\% and 48\% of yearly numbers during the declining phases of SC23 (rising-phase years: 1997--1999; declining-phase years: 2003--2008) and SC24 (rising-phase years: 2009--2010; declining-phase years: 2015--2017), respectively. In fact, as was noted by \citet{Echer2013}, SIR/HSS events were responsible for 30\% of the moderate geomagnetic storms taking place during the rising phase of SC23.

\begin{table}
 \caption{Subdivision of solar cycles (SC) between 1995 and 2017}
 \label{table1}
 \centering
 \begin{tabular}{l l l c}
 \hline
 \textbf{Phase} & \textbf{Start date} & \textbf{End date} & \textbf{Duration (months)}   \\
 \hline
 SC 22 -- late declining  & Jan 1995 & Jul 1996 & 19 \\
 SC 23 -- rising          & Aug 1996 & Dec 1999 & 41 \\
 SC 23 -- maximum         & Jan 2000 & Jun 2002 & 30 \\
 SC 23 -- early declining & Jul 2002 & Dec 2005 & 42 \\
 SC 23 -- late declining  & Jan 2006 & Nov 2008 & 35 \\
 SC 24 -- rising          & Dec 2008 & Aug 2011 & 33 \\
 SC 24 -- maximum         & Sep 2011 & Jun 2014 & 34 \\
 SC 24 -- early declining & Jul 2014 & Dec 2017 & 42 \\
 \hline
 \end{tabular}
 \end{table}
 
The following common characteristics appear to repeat during the two observed cycles. After the sunspot minimum is reached, the number of SIR/HSS events is at minimum in the beginning of a new cycle both during cycles 23 (year 1997) and 24 (year 2009), which is consistent with findings reported by \citet{Jian2011}. Then the number of SIR/HSSs starts to increase during the next two years in the rising phase. During sunspot maximum years the increase stops and there is a local minimum (year 2000 for cycle 23 and year 2014 for cycle 24). Even if one added the HSSs rejected because they contain an embedded ICME (part of the orange bars in Figure~\ref{fig3}), the number of SIR/HSS events would still drop and a local minimum would still be found during these years (2000 and 2014). This confirms the observation made by \citet{Jian2006} during the maximum phase of solar cycle~23.

In the early declining phase of solar cycles, the number of SIR/HSS events starts to increase again and the maximum number of events occurs obviously during the last year of cycle 22 (1996, but we have data only for the last two years) and the second last year (2007) for cycle 23. For cycle 24, the number of events has almost the same values for the last two observed years, i.e., 2016 and 2017. If the length of cycle 24 were the nominal 11 years, this would imply that sunspot minimum would take place in 2019 and we could expect to have the maximum number of SIR/HSS events in year 2018 or 2019. The number of SIR/HSS events in 2007 is 39, and there were 35 events in years 1996 and 2016. These numbers correspond roughly to 3 SIR/HSS events per month with maximum speeds exceeding 500 km/s during maximum occurrence.

\begin{figure}
 \centering
 \includegraphics[width=.8\textwidth]{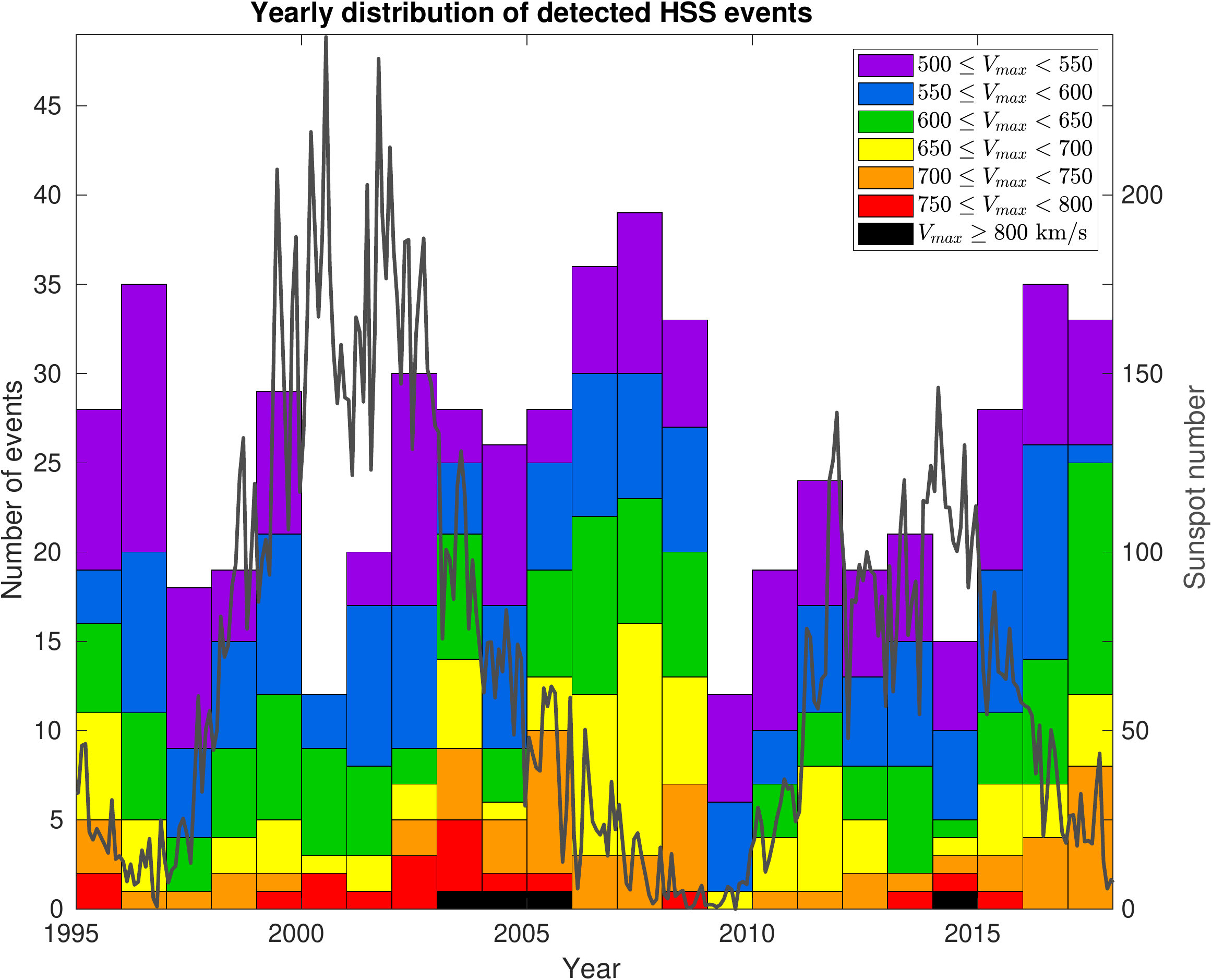}
 \caption{Histogram: Yearly distribution of high-speed stream events detected with the method and binned according to their maximum solar wind speed value. Line plot: Monthly sunspot number.}
 \label{fig6}
\end{figure}

One interesting characteristic of the detected SIR/HSS events is the distribution of their maximum solar wind speed. Figure~\ref{fig6} gives the yearly distribution of the 588~detected SIR/HSS events, binned according to their maximum solar wind speed value during the first three days in the 1~h resolution data. The following clear features can be observed. First, during the first full year of a new cycle, when the number of SIR/HSS events is very low, the velocities of the HSSs are also low (years 1997 and 2009) with maximum velocities remaining mainly below 600~km/s. Second, during the declining phases, when the number of SIR/HSS event maximizes, the proportion of events with velocities higher than 650~km/s also increases. Third, HSS velocities larger than 750~km/s are very rare and occur typically 0-2 times per year. Year 2003 is an exception with five events having maximum velocities exceeding 750~km/s (starting on 26~July, 5~August, 19~August, 16~September, 7~December).

\subsection{SIR/HSS Event Evolution as a Function of Solar Cycle Phases}

\begin{figure}
 \centering
 \includegraphics[width=\textwidth]{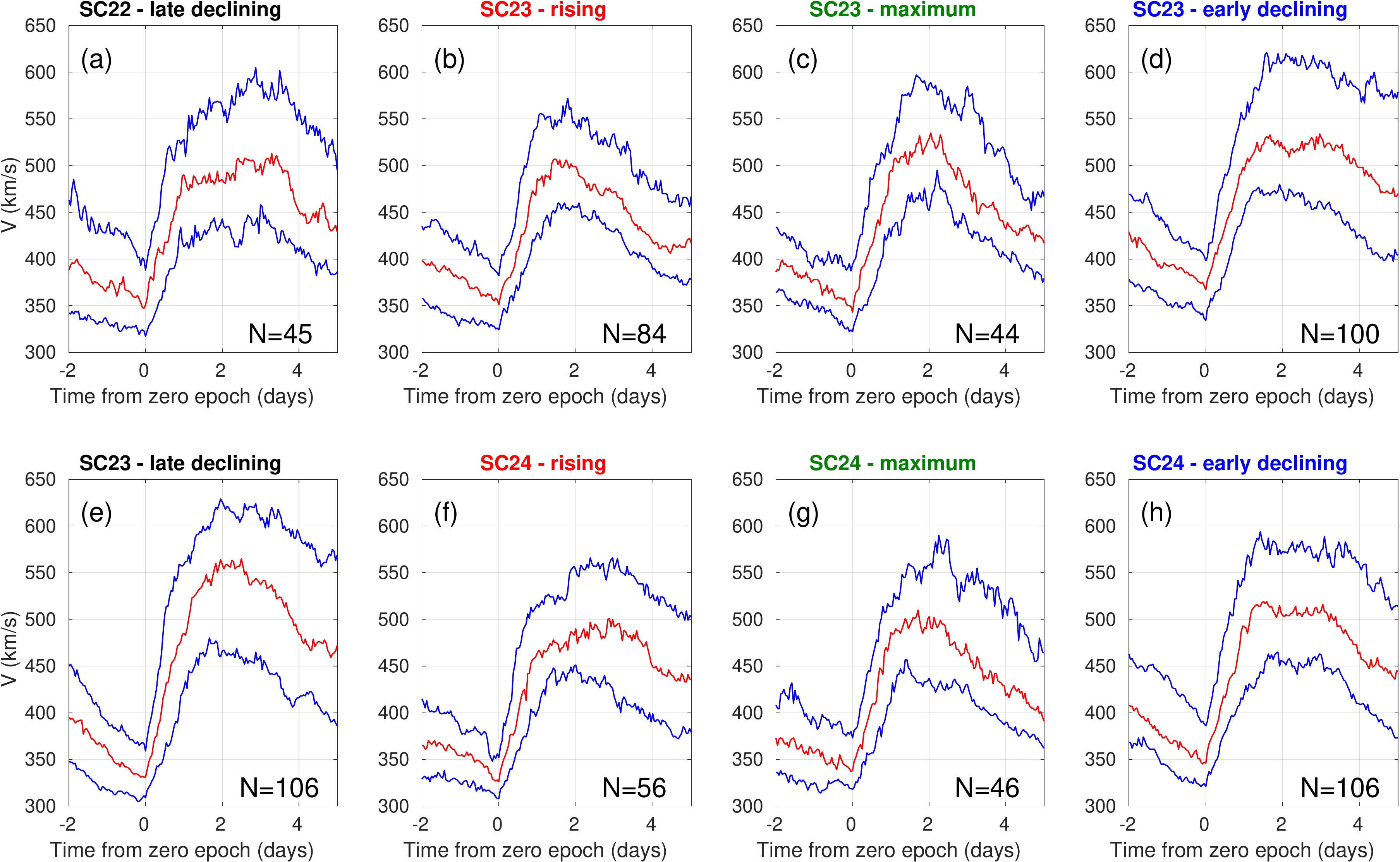}
 \caption{Superposed epoch analysis of the solar wind speed during SIR/HSS events during (a)~the late declining phase of SC22, (b)~the rising phase of SC23, (c)~the maximum of SC 23, (d)~the early declining phase of SC23, (e)~the late declining phase of SC23, (f)~the rising phase of SC24, (g)~the maximum of SC24, and (h)~the early declining phase of SC24. The number indicated in the bottom-right-hand corner of each panel corresponds to the number of SIR/HSS events during the corresponding solar cycle subdivision (cf. Figure~\ref{fig5}).}
 \label{fig7}
\end{figure}

\begin{figure}
 \centering
 \includegraphics[width=\textwidth]{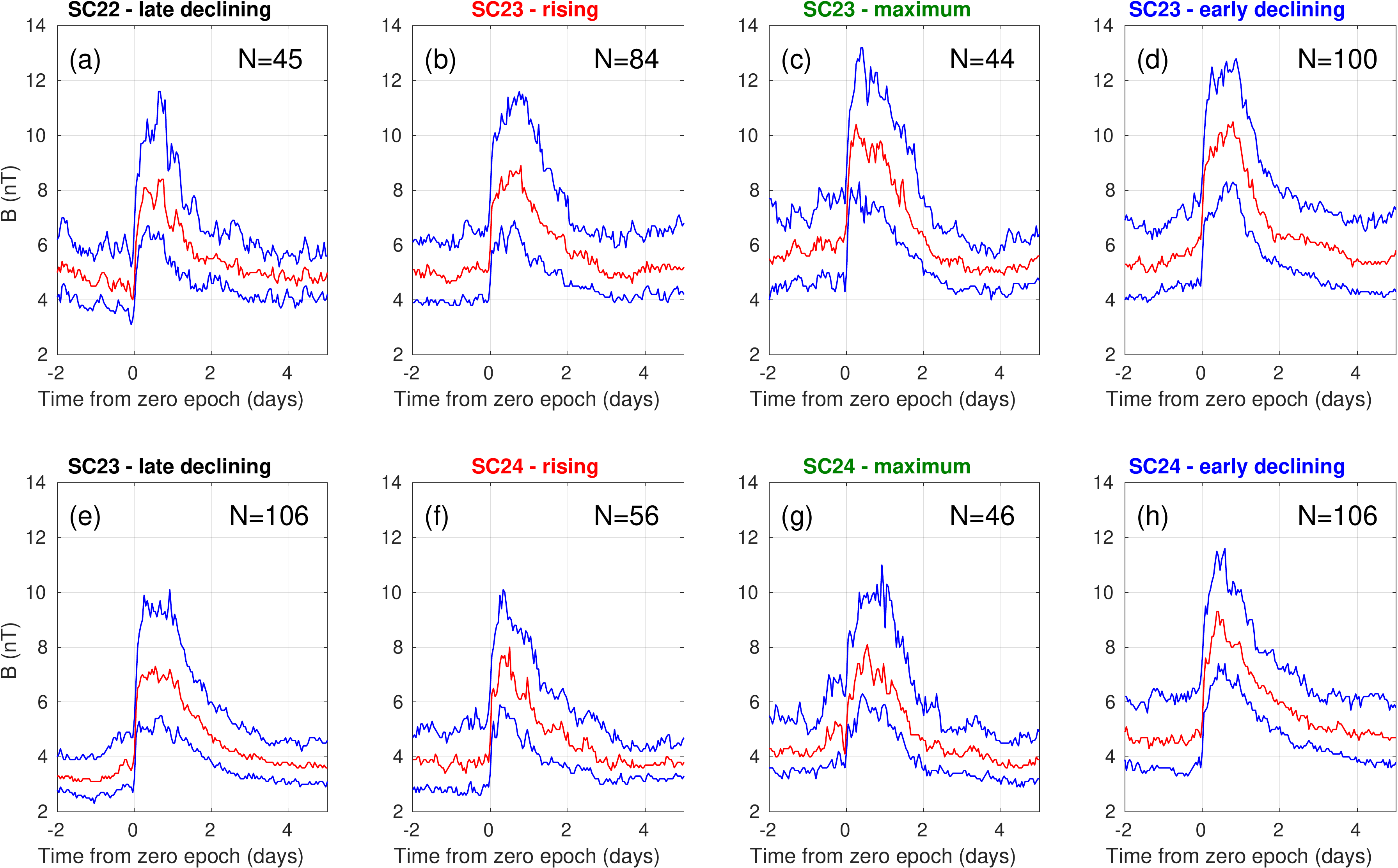}
 \caption{Same as Figure~\ref{fig7} for the IMF magnitude.}
 \label{fig8}
\end{figure}

\begin{figure}
 \centering
 \includegraphics[width=\textwidth]{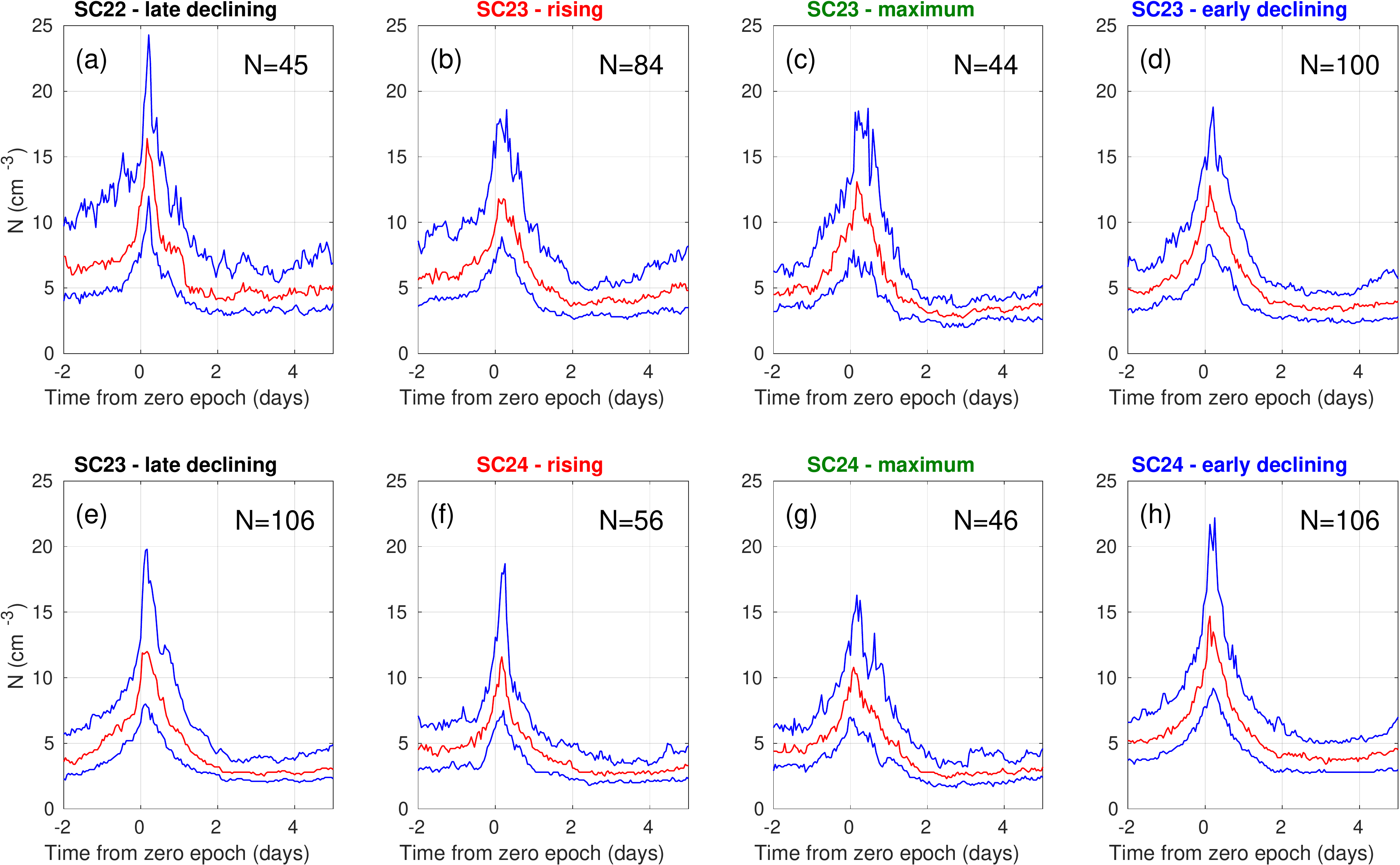}
 \caption{Same as Figure~\ref{fig7} for the solar wind density.}
 \label{fig9}
\end{figure}

\begin{figure}
 \centering
 \includegraphics[width=\textwidth]{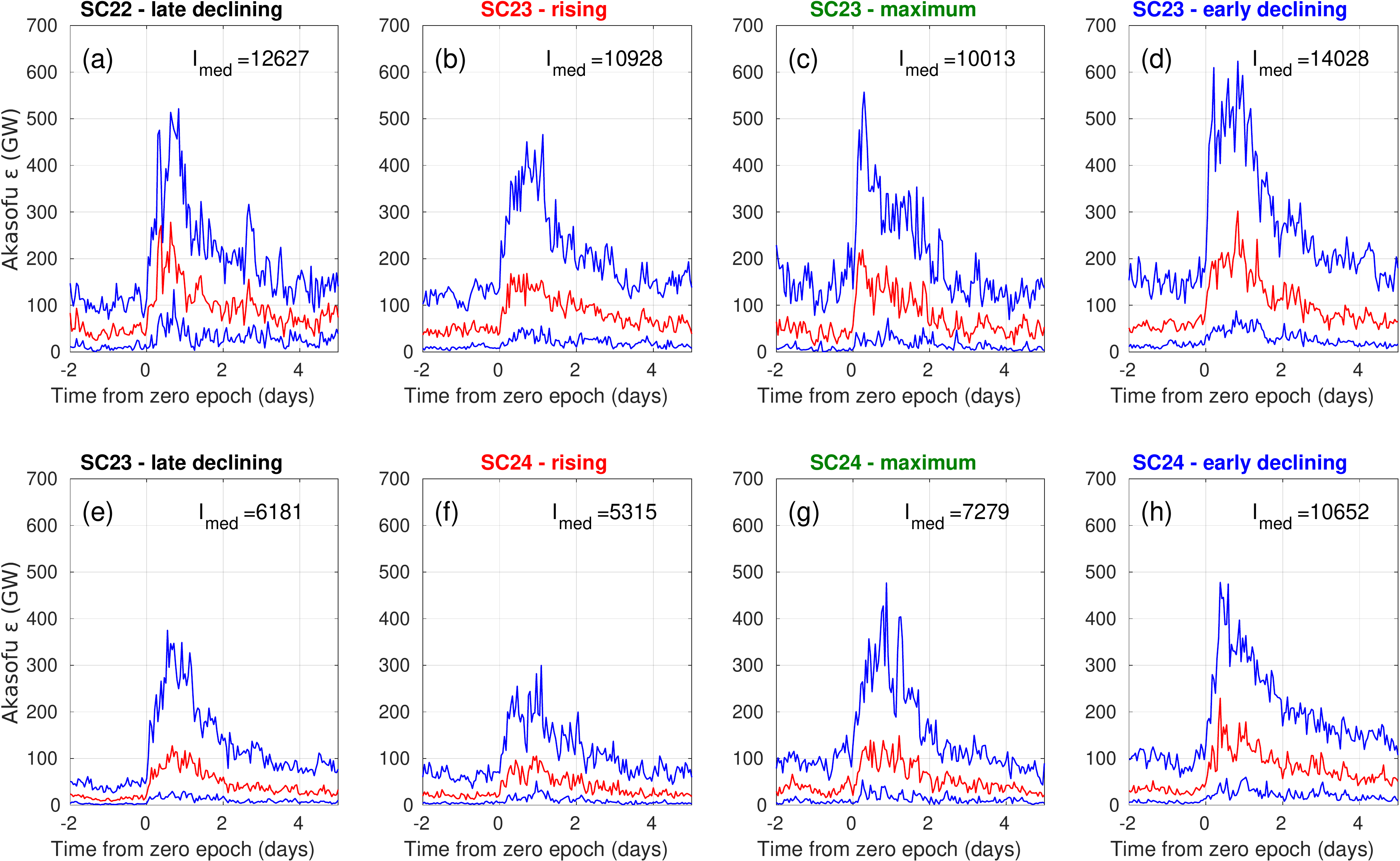}
 \caption{Same as Figure~\ref{fig7} for the Akasofu $\varepsilon$ parameter, except that the number given in the top-right-hand corner of each panel corresponds to the integrated value of the median curve from the zero epoch until day~5 (in GWh).}
 \label{fig10}
\end{figure}

\begin{figure}
 \centering
 \includegraphics[width=\textwidth]{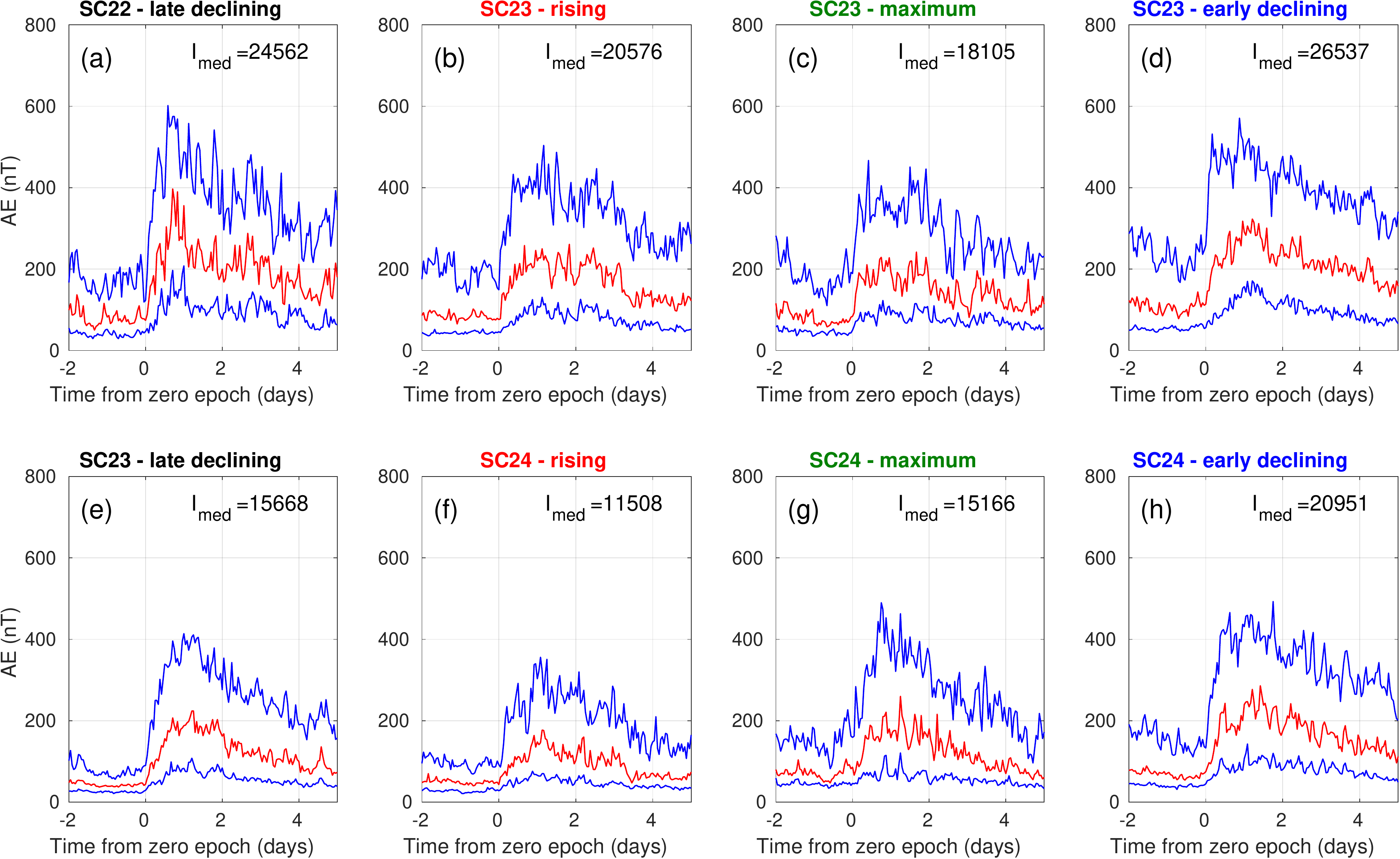}
 \caption{Same as Figure~\ref{fig10} for the $AE$ index (integrated values in nTh).}
 \label{fig11}
\end{figure}

\begin{figure}
 \centering
 \includegraphics[width=\textwidth]{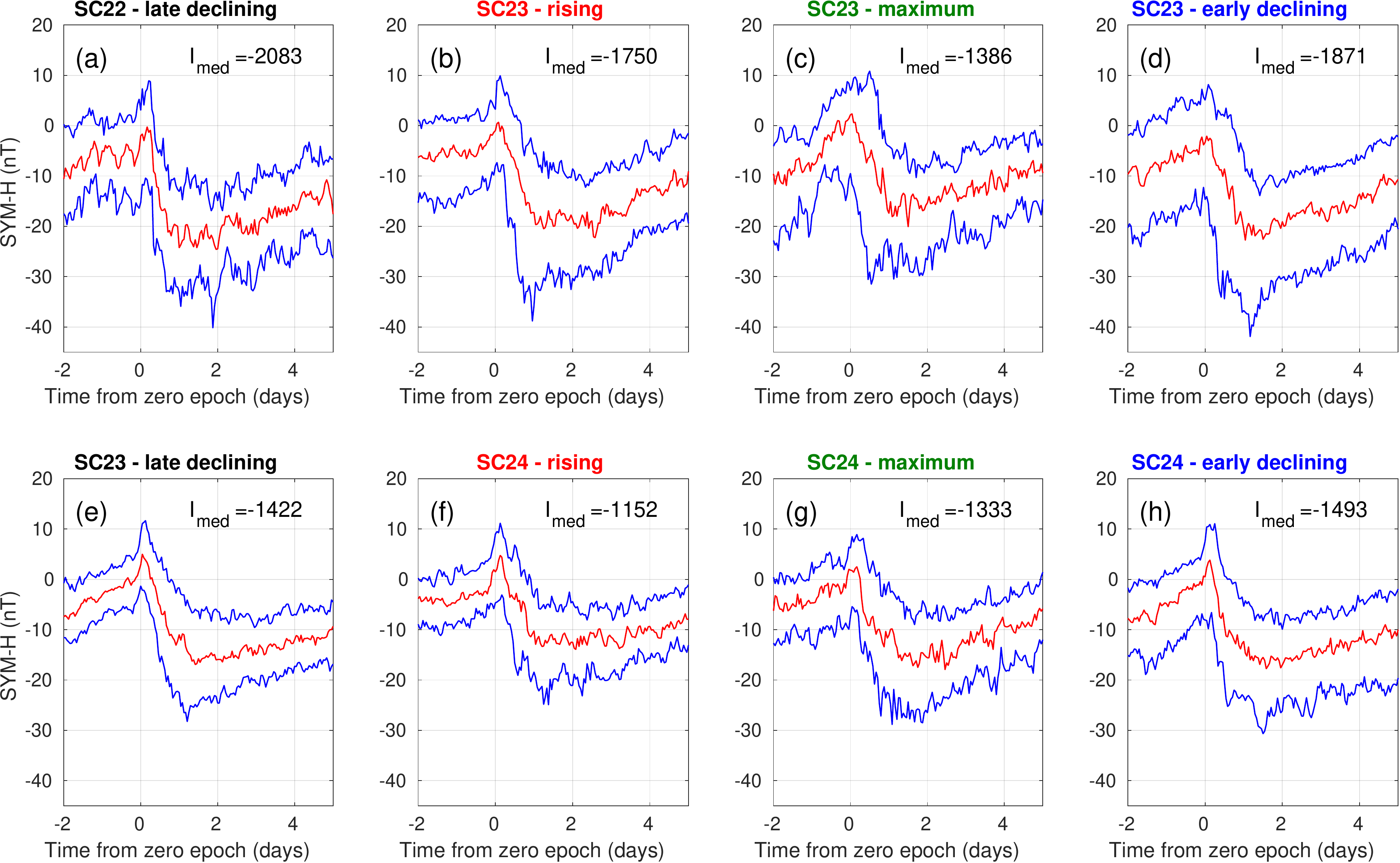}
 \caption{Same as Figure~\ref{fig10} for the \emph{SYM--H} index (integrated values in nTh).}
 \label{fig12}
\end{figure}

A comparison of SIR/HSS features during the solar cycle phases defined in Figure~\ref{fig5} is made using the superposed epoch analysis method, which was used in \citet{Grandin2015,Grandin2017}. The superposed epoch method is a statistical analysis invented by \citet{Chree1913}, based on the collection of a large number of events, the definition of a reference time called ``zero epoch'' in the event timeline and the extraction of statistical properties (e.g., mean, median, and standard deviation) of the physical parameters of the events as a function of time relative to the zero epoch \citep[see also 3.1 in][]{Grandin_PhDthesis}. Figures~\ref{fig7} to \ref{fig12} show the statistical behaviors of key solar wind parameters and geomagnetic indices during high-speed stream events taking place at each solar cycle phase. 

Figure~\ref{fig7} shows the superposed epoch analysis of the solar wind velocity during the SIR/HSS events detected during (a) the late declining phase of SC22, the (b) rising, (c) maximum, (d) early declining and (e) late declining phases of SC23, and the (f) rising, (g) maximum and (h) early declining phases of SC24. In each panel, median values are shown in red, and upper and lower quartiles in blue. The data are plotted from two days before the zero epoch until five days after, the zero epoch of each event corresponding to its starting time given by the algorithm. The number of events used to obtain the curves in each panel is given in the bottom-right-hand corner of the panel.  

Out of these eight panels, the highest median velocities take place during the late declining phase of SC23 covering years 2006--2008 (panel e). The maximum median velocity in these superposed epoch plots reaches values of $\ge$550~m/s from the end of day~2 to the beginning of day~3 (please note that values 0--1 in the $x$ axis correspond to day~1 and values from 1 to 2 to day~2). Second largest velocities are found in the early declining phases of SC23 and SC24 (panels d and h). Clearly smallest velocities occur in the rising phases of solar cycles (panels b and f).

It looks a bit surprising that the velocity in the late declining phase of SC22 is not showing particularly high values, unlike in SC23. The answer may be related to the special characteristics of the long-lasting solar minimum at the end of SC23. The study of coronal holes by \citet{deToma2011} provides insight in this topic. According to \citet{deToma2011}, during the 1996 solar minimum, the two large polar coronal holes were the dominant coronal holes on the Sun and stable sources of fast solar wind over the poles, like in typical solar minimum conditions. This gave a relatively slow solar wind at the Earth, originating mostly from the edges of the polar coronal holes. However, the long period of low solar activity from 2006 to 2009 was characterized by weak polar magnetic fields and polar coronal holes smaller than observed during the previous minimum. Instead, large low-latitude coronal holes were present on the Sun until 2008 and remained important sources of recurrent high-speed solar wind streams. By early 2009, most the low-latitude coronal holes had closed down. With the increase of the new SC24 activity, small, mid-latitude coronal holes appeared. The fast wind in the rising phase was coming mostly from the edges of the polar coronal holes and occasionally from the small, mid-latitude coronal holes. Finally, the high-speed wind in the declining phases comes typically from the low-latitude coronal holes \citep{deToma2011}.

Figure~\ref{fig8} shows the IMF magnitude $B$ in the same format. The largest values for $B$ in the SIR are found in the early declining and maximum phases of SC23 with the median values exceeding 10~nT during day 1. Inside SC24, the largest values of $B$ take place in the early declining phase with median values reaching 9~nT. This is a confirmation that the maximum $B$ value in SIRs does not simply follow solar activity given by the monthly sunspot number, as was observed by \citet{Jian2006} when studying SC23. The lowest values are observed in the late declining phase of SC23, when the peak values of the median curve just reach 7~nT. These values are clearly smaller than in the late declining phase of the previous cycle, SC22. However, the difference comes mainly from the fact that the magnetic field values prior to the arrival of the SIR are clearly smaller in SC23, namely about 3~nT on {day~$-1$}, whereas in SC22 the corresponding value is about 4.5~nT. The low values of $B$ at 1~AU near the ecliptic plane during the 2007--2008 solar minimum period of SC23 have been a topic of many studies \citep[see, e.g.,][]{Lee2009,Wu2013}.

When comparing SC23 and SC24, a clear overall difference can be found. It appears that the SC23 data (panels b, c, and d) have their peaks of median $B$ 15--20\% higher than the corresponding SC24 data (panels f, g, and h).

Figure~\ref{fig9} shows the superposed epoch analysis of the solar wind number density. The density has a narrow peak very shortly after the zero epoch for each solar cycle phase. The largest median values are obtained in the late declining phase of SC22 and the early declining phase of SC24, in both cases with values close to 15~cm$^{-3}$. Again, the late declining phase of SC23 is anomalous, since the background densities before the arrival of SIRs are only about 3.5~cm$^{-3}$ in SC23, while in SC22 the values are about 6~cm$^{-3}$. In the region of high speeds, the solar wind is dilute, and the densities are about 2.5~cm$^{-3}$ in SC23, while in SC22 the densities are about 4~cm$^{-3}$. \citet{Lee2009} noticed that when all solar wind OMNI data are considered, the peak occurrence for SC22 is centered at 3.5~cm$^{-3}$ and for SC23 at 2.5~cm$^{-3}$. Obviously the rarefaction regions of HSSs dominate in the overall OMNI data.

Figure~\ref{fig10} shows the superposed epoch analysis of Akasofu~$\varepsilon$ (expressed in GWh), which has been calculated by using equation~(1). Since the $\varepsilon$ parameter depends linearly on solar wind velocity $V$ and on the quadratic value of $B$, we expect that the cycle phases where these parameters maximize will have the largest values for $\varepsilon$. In addition, there is a strong dependence on the IMF clock angle, which is not shown in the previous plots. However, across several tens of events, the clock angle can be expected to take very different values, as the IMF will be northward in some events, southward in others, and with $B_z \simeq 0$ in yet some other events. Therefore, one may expect that, in the superposed epoch analysis, the effect of the clock angle on the value of Akasofu $\varepsilon$ is averaged out. It can be seen that the highest individual median value of the $\varepsilon$ parameter takes place in the early declining phase of SC23, where $B$ has the largest values and $V$ is relatively large. Almost as large values are seen in the late declining phase of SC22, which may be explained by the fact that the solar wind speed reached high values while $B$ was still peaking.

In order to assess the sustained impact of the SIR/HSSs on the geospace environment, the median values of the Akasofu $\varepsilon$ parameter were also integrated from the zero epoch until day~5, and those values are given in the panels of Figure~\ref{fig10} as $I_\text{med}$. The early declining phase of SC23 shows the largest integrated value of Akasofu~$\varepsilon$. It also has the largest upper quartile values. The second highest integrated value is in the late declining phase of SC22, and the third highest value in the early declining phase of SC24. Even though the late declining phase of SC23 had the largest velocities, the Akasofu $\varepsilon$ parameter has low values, due to the low values of $B$ during the extended solar minimum, as discussed above. These are the second lowest values of all the studied phases. The lowest values are found in the rising phase of SC24.

Overall, Akasofu~$\varepsilon$ remains significantly lower during SC24 compared to SC23. In each phase, the energy input peaks around day~1 following the zero epoch, the value remains high until the end of day 2 and then decreases but remains somewhat elevated at least until day 5. 

Figures~\ref{fig11} and \ref{fig12} show the $AE$ and \emph{SYM--H} indices, respectively. In general, the maximum of substorm ($AE$) and storm (\emph{SYM--H}) activity is reached about one day after the zero epoch, in the beginning of day~2, which is before the solar wind speed reaches its maximum and hence corresponds to the stream interaction region rather than the HSS itself. \citet{Burlaga1977} showed that SIRs often amplify Alfv{\'e}n waves propagating away from the Sun, which cause the IMF $B_z$ component to alternate between northward and southward and therefore produce intermittent substorm activity while the SIR passes by Earth. Alfv{\'e}n waves are also often present throughout the fast streams, which are also geoeffective. This is illustrated by the fact that the $AE$ index remains enhanced at least until day~5. It is well-known that the geoeffectiveness of HSSs in terms of substorm activity is also largely controlled by the sign of $B_z$ \citep[e.g.][]{Kavanagh2012}, which explains the rather large dispersion of the $AE$ index values following the zero epoch indicated by the upper and lower quartile values in blue and the fairly broad $AE$ peak.

Akasofu~$\varepsilon$ orders the $AE$ index values quite well, so that the three most intense phases and the weakest one (rising phase of SC24) are the same for $AE$ as for Akasofu~$\varepsilon$. The same holds for \emph{SYM--H}, except that the order of the two most intense events is reversed. Typically, the storms produced by SIR/HSSs are rather weak and only a few percent of the SIRs can produce intense geomagnetic storms ($Dst<-100$~nT) \citep{Richardson2006}. However, HSSs are quite effective in producing substorms, and these events are called HILDCAA (high-intensity long-duration auroral activity) by \citet{Tsurutani2006}. It was found by \citet{Hajra2013} that HILDCAAs are on average 20\% longer during the declining phase than during the rising and maximum phases. Figure~\ref{fig11} shows indeed that during the declining phases the $AE$ index is still elevated on day~5. In addition, the maximum $AE$ values are higher during the early declining phases of SC23 and SC24 than during the other phases.  

Again, the top-row panels of Figures~\ref{fig11} and \ref{fig12} exhibit a more intense response than the bottom-row panels, underlining that geomagnetic activity related to SIR/HSSs significantly decreased between the early and late declining phases of SC23. Within a given solar cycle, the rising and maximum phases are those with lowest geomagnetic activity during SIR/HSS events, whereas the early and late declining phases exhibit stronger activity. 

\begin{figure}
 \centering
 \includegraphics[width=\textwidth]{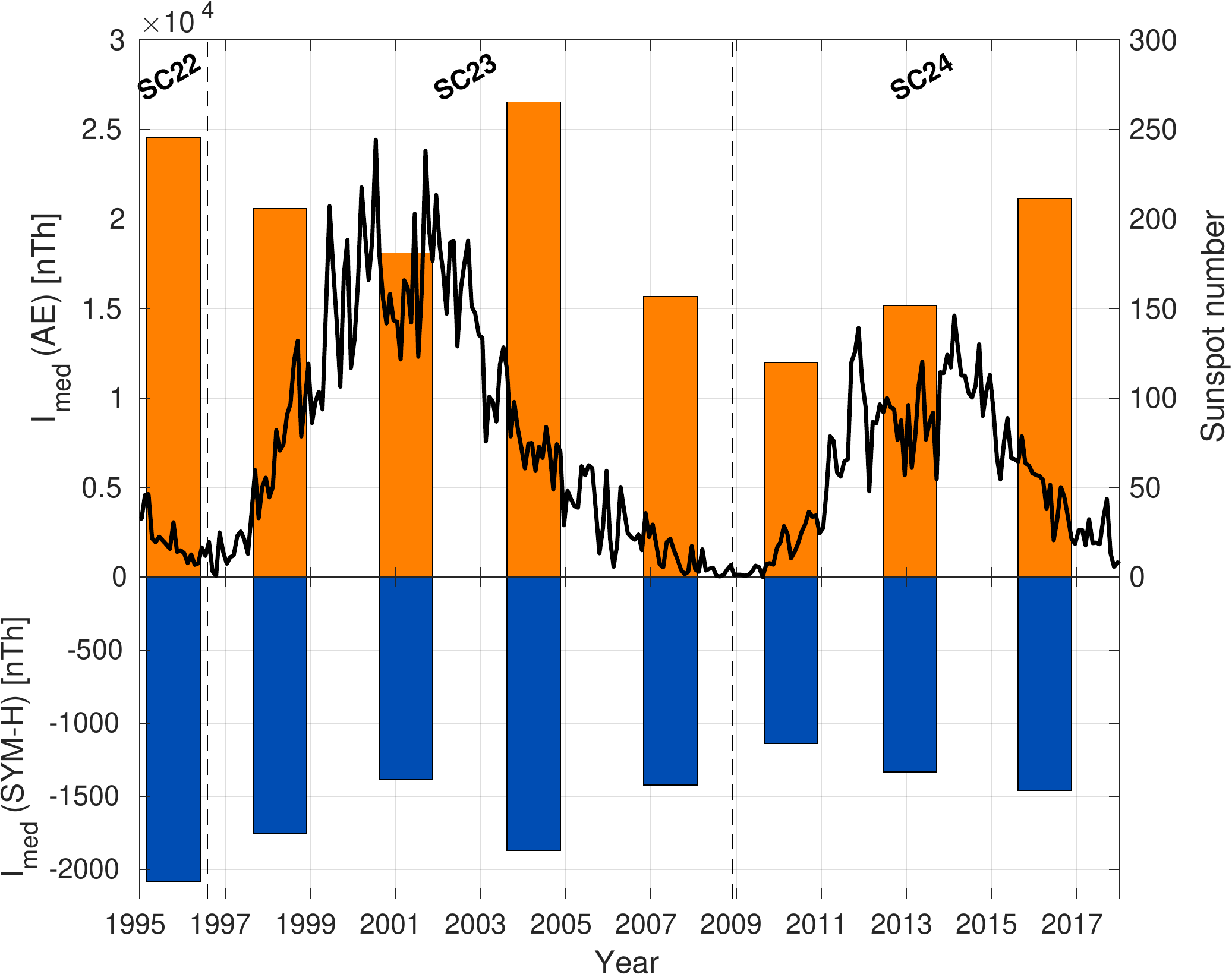}
 \caption{Bar diagram: Integrated values of median curves from the zero epoch until day~5 for the $AE$ (top) and $SYM$--$H$ (bottom) indices (see Figures~\ref{fig11} and \ref{fig12}). Line plot: Monthly sunspot number.}
 \label{fig13}
\end{figure}

Figure~\ref{fig13} visualizes the 5-day integrated values of geomagnetic activity measured by $AE$ (orange bars) and $SYM$--$H$ (blue bars) during SIR/HSS events within the defined solar cycle phases from 1995 to 2017, superposed on the monthly sunspot number (black line). This figure shows the well-known importance of declining phases of the solar cycles for geomagnetic activity. For SC22, we have data only for the late declining phase, and it is the most geoeffective one in terms of $SYM$--$H$, and the second-most geoeffective in terms of $AE$ for all data. Inside SC23, the early declining phase is the most geoeffective one in terms of both indices, and in terms of $AE$ the most geoeffective one of all data. The anomalous late declining phase of SC23 is the least geoeffective one, and even the rising phase of this cycle is more geoeffective. The maximum phase is less geoeffective than the rising phase in SC23. When moving to SC24, the geoeffectiveness clearly decreases compared to SC23, as does the sunspot number. The early declining phase is the most geoeffective one inside SC24, but we do not yet have data for the late declining phase. Contrary to SC23, the maximum phase is more geoeffective than the rising phase. Comparing SC24 to SC23 in terms of auroral electrojet and substorm activity measured by the $AE$ index, the geoeffectiveness has been about 40\% lower in the rising phase, 15\% lower in the maximum phase, and 20\% lower in the early declining phase. For the ring current and storms measured by the $SYM$--$H$ index, the geoeffectiveness has been about 35\% lower in the rising phase and 20\% lower in the early declining phase, but roughly similar during the maximum phase.

As a summary, we find that the geoeffectiveness of SIR/HSSs varies in the different phases of the solar cycle. Our results are in line with the often presented statement that SIR/HSSs are the most geoeffective during the declining phases of solar cycle \citep[e.g.,][]{Hajra2013,Kilpua2017,Chi2018}, even though we find that the late declining phase of SC23 was very weakly geoeffective owing to reasons discussed in the next section.

\section{Discussion}
 
We will discuss two major topics. First, the validity of the detection algorithm is assessed by comparing the obtained list of SIR/HSS events with existing catalogs during years when they overlap. Second, the identified solar cycle phase-dependent features of SIR/HSSs and their geoeffectiveness will be considered in the global picture of the known features of the studied solar cycles.

\subsection{Comparison with other HSS event lists}

We compared the events detected by our algorithm with the HSS catalog issued by Mari\cb{s} Muntean et al. (\url{http://www.geodin.ro/varsiti/}), which covers years from 2009 until 2016. We selected year 2015 for the comparison, as it is at the beginning of the declining phase of SC24 and hence contains a mix of HSSs, ICMEs and slow solar wind. The Mari\cb{s} Muntean et al. catalog is solely based on the solar wind speed, averaged at 3~h resolution. A HSS event is detected if the maximum 3~h mean speed during one day exceeds the minimum 3~h mean speed of the previous day by at least 100~km/s, and if the enhanced speed persists for two days. In 2015, with their method, Mari\cb{s} Muntean et al. detected 44~HSS events. Comparatively, our algorithm detected only 27~SIR/HSS events during that year. Two main reasons explain this first difference: (i) the Mari\cb{s} Muntean et al. catalog does not focus on SIR-related HSSs, and hence contains 11~events corresponding to (or contaminated by) ICMEs (by comparison with the Richardson and Cane list), and (ii) the Mari\cb{s} Muntean catalog is not restricted to HSSs with $V_\text{max} \geq 500$~km/s, and therefore contains another 10~events whose maximum velocities lie between 400 and 500~km/s. On the other hand, our algorithm detected four SIR/HSS events which are not in the Mari\cb{s} Muntean et al. catalog (on 14~July, 18~August, 2~September, and 28~November 2015). Figure~S1 in Supporting Information illustrates the comparison of the two lists for 2015. For the 23~events common to both catalogs, 16 of them have their starting time with less than 6~h difference, and only one of them has starting time differing by more than 12~hours (2~January, 23~UT vs 4~January, 9--12~UT). For this specific event, our starting date nicely corresponds to the solar wind density peak, which makes this detection time consistent with the rest of the list (see Figure~9).

Another HSS catalog was earlier compiled by \citet{Gupta2010} for years 1996--2007. The retained criteria are essentially the same as those used by Mari\cb{s} Muntean et al., i.e., an increase in the solar wind speed of at least 100~km/s with sustained velocity for at least two days. Year 1996 was chosen for comparison of our list with the \citet{Gupta2010} catalog, since it corresponds to the peak of SIR/HSS activity at the transition between SC22 and SC23. \citet{Gupta2010} listed 40~HSS events in 1996, but several of them actually contained two to three distinct streams, therefore representing at total of 58~individual streams (compared to 35~events from our method). Among these, 22~events were associated with solar wind speeds remaining below our threshold of 500~km/s. One of their events (19~September) was not detected by our algorithm, since its initial solar wind speed was greater than 450~km/s (see criterion~2a), and another one (the second stream of the event starting on 8~April) was also rejected for being too soon after our 11~April event (see criterion~3). On the other hand, our algorithm detected a SIR/HSS event on 2~July, which is not included in the \citet{Gupta2010} catalog. Besides these few differences, there is an overall good agreement between both catalogs. Figure~S2 in Supporting Information illustrates the comparison of the two lists for 1996.

A third list of HSSs which may be used for comparison is the one given in \citet{Morley2010}, which uses different criteria for SIR/HSS identification. As a first criterion, \citet{Morley2010} looked for an enhancement in the radial velocity of the solar wind ($V_x$) associated with a west--east deflection of the flow, i.e., a change in the sign of $V_y$. For the obtained SIR/HSS candidates, other solar wind parameters such as the IMF magnitude, the proton number density and the proton temperature were then examined to reject false positives. In addition, two consecutive events had to be separated by at least 2~days, and events exhibiting clear ICME signatures were removed from the list. The list given in Table~2 of \citet{Morley2010} considers 67~stream interfaces verifying those criteria between 2005 and 2008. During these same years, our algorithm detected 136~SIR/HSS events, which is more than twice as many. Yet, of the 67 events studied by \citet{Morley2010}, we found that nine of them were not detected by our algorithm. In two of these cases, a ICME was reported in the Richardson and Cane list (criterion~4), in three cases, the maximum solar wind speed did not reach 500~km/s (criterion~2c), and in four cases a previous event was detected less than three days before the rejected event (criterion~3). Figures~S3--S6 in Supporting Information illustrate the comparison of the two lists for 2005--2008.

For the 58~SIR/HSS events in common with the \citet{Morley2010} list, a comparison of the starting times reveals that in all cases but one our algorithm gives an earlier time. This is inherent to the difference in the criteria for setting the beginning of the events: while our algorithm retains the time after which the solar wind speed starts increasing, \citet{Morley2010} were interested in the stream interface itself, which is determined with the azimuthal flow reversal ($V_y$ sign change). The difference $\Delta t_0$ between those starting times can vary significantly. Figure~S7 in Supporting Information gives a histogram of $\Delta t_0$, which is defined as positive when our algorithm gives an earlier time than in the \citet{Morley2010} list. The peak in the distribution of $\Delta t_0$ is for values comprised between 6 and 12~hours, but up to 70~hour difference was found for a couple of events. The first such event (3 May 2006) exhibits steady solar wind speed enhancement for three days before the stream interface, and in the second case (7 February 2008), our method detects a small stream immediately preceding the main event, hence causing an erroneously early starting time. The difference is negative in only one case (22 April 2008), for which the start time given by our algorithm corresponds nicely to the beginning of the solar wind speed increase.

Finally, the list obtained with the detection method presented in this paper can be compared to the catalog compiled by \citet{Jian2006}, later extended to cover years 1995 to 2009 (extended version available at \url{http://www-ssc.igpp.ucla.edu/~jlan/ACE/Level3/SIR_List_from_Lan_Jian.pdf}). Between 1995 and 2009, the \citet{Jian2006} list contains 437~SIR/HSSs with $V_\mathrm{max} \geq 500$~km/s, whereas our list contains 394~events during the same time 15-year period. When making a detailed analysis of the differences between those two lists, it appears that 39~events from our list are not present in the Jian et al. list. These events have each been examined visually to ensure they do exhibit typical SIR/HSS signatures. The reason for the difference may be that the event identification by \citet{Jian2006} imposes that the events must exhibit at least five out of the seven expected features considered in their study, which are solar wind speed increase, total perpendicular pressure pileup, velocity deflections, proton density enhancement, proton temperature enhancement, increase in entropy, and magnetic field compression. On the other hand, 82~events with $V_\mathrm{max} \geq 500$~km/s in the Jian et al. list are not present in ours. The reasons are: presence of an ICME with $V_\mathrm{max} \geq 500$~km/s within [$t_0-1$~day, $t_0+3$~days] (32~events); $V_\mathrm{max} < 500$~km/s in the hourly OMNI data whereas the Wind or ACE data with 93~s and 64~s resolution, respectively, used by Jian et al. exceeds 500~km/s during the event (25~events); solar wind speed greater than 450~km/s at the start time of the event (15~events); compound events separated by less than 3~days (9~events); no sharp IMF magnitude gradient (1~event). Besides, in 27~cases, event starting dates differed by more than 24~h from one list to the other. About half of these cases correspond to compound events for which in one list the first event only is mentioned, while in the other list the second event only is mentioned. In the other cases, our list gives an earlier time than the Jian et al. list because the detected IMF magnitude enhancement is earlier than the leading edge of the SIR as identified by Jian et al. Overall, the agreement between both lists over those 15~years of comparison is satisfactory, and all the differences can be explained by the choices made in the criteria for identification in this study versus the Jian et al. study.

While some differences may be noted with each of those above-discussed catalogs, one advantage of the list obtained with this algorithm is that it contains SIR/HSS events during 23 years, i.e., roughly two solar cycles. Provided an updated list of ICMEs reaching the Earth exists and solar wind observations remain available, this list of SIR/HSS events can be completed during the upcoming years by applying the same algorithm, thus enabling further studies about the upcoming solar cycles.

\subsection{Solar cycle variability of SIR/HSSs and geoeffectiveness}

We have briefly discussed in the previous sections the properties of the solar wind at 1~AU near the ecliptic plane during the extended solar minimum of SC23 observed in other studies. Comparing the minimum periods of SC22 and SC23, \citet{Lee2009} noticed the following. For the IMF magnitude, the peak of the distribution for the SC23 period is centered at 3.5~nT, which is 30\% less than 5~nT, the approximate central value for the peak of the SC22 distribution. For the number density, the distribution is shifted toward lower values by about 30\%. For solar wind velocity, both the SC22 and SC23 velocity distributions have peak values occurring around 340~km/s, but for SC23 the peak is lower and the high-speed tail distribution, which is centered near 580~km/s, has slightly larger percent occurrence during SC23.

In this study, we have specifically studied the SIR/HSS events and our findings are in general accordance with \citet{Lee2009}, but give some additional insight into the distribution functions. We found by comparing the late declining phases of SC22 and SC23, namely years 1995--1996 and 2006--2008, the following properties. Both the IMF magnitude and density values before the SIR/HSS events are about 30\% lower in SC23 than SC22, but the compressions of $B$ and $N$ are equally strong in both cycles. The peak of the density distribution in \citet{Lee2009} study comes obviously from the dilute solar wind in the high-speed stream region (in our figures superposed epoch after day~2).

Comparison of the distributions of HSS peak velocities during the late declining phases of SC22 and SC23 shown in Figure~\ref{fig6} indicates that the proportion of $500<V_{max}<600$~km/s is higher in SC22, whereas the proportion of $V_{max}>700$~km/s is higher in SC23. The superposed epoch presentation of velocities shown in Figure~\ref{fig7} confirms that the median peak velocities are higher in SC23 ($V_{max}\sim 560$~km/s) than SC22 ($V_{max}\sim 510$~km/s). These are in harmony with the observations of the behavior of the high-speed tail distributions by \citet{Lee2009}.

The solar wind parameters control the coupling of the IMF with the terrestrial magnetosphere and energy transfer from the solar wind into the magnetosphere--ionosphere system. Several coupling functions have been developed \citep[see, e.g.,][]{Newell2007}, which all depend on solar wind velocity and magnetic field. In this study, we have used the traditional Akasofu~$\varepsilon$. We saw a very similar behavior between the Akasofu~$\varepsilon$ and the $AE$ and $SYM$--$H$ indices. This can be expected, since it is well-known that the solar wind parameters directly affect the geoeffectiveness of SIR/HSSs. The surprisingly poor geoeffectiveness of the SIR/HSS events in the late declining phase of SC23 is obviously explained by the unusually low magnetic field values during the deep minimum of SC23 discussed above. Even though the velocities of these HSSs were higher than in the late declining phase of SC22, this did not make those more geoeffective, and this can be explained by the quadratic dependence on the magnetic field intensity compared to the linear dependence on the velocity in the coupling function. The fact that SIR/HSSs are less geoeffective during SC24 than during SC23 also finds an explanation in the solar wind properties. The velocities were a bit smaller during SC24 than SC23, but the magnetic fields were even more significantly smaller. This will be discussed more below.

Following the extremely low minimum of SC23, SC24 has shown unusually low activity \citep{Kamide2013}. In particular, \citet{McComas2013} showed that the rising phase of SC24 exhibited significantly lower key parameters (e.g., IMF magnitude, solar wind proton density, temperature, velocity, dynamic pressure) than the rising phases of previous solar cycles, which correlates with the fact that SC24 has been the weakest solar cycle since the beginning of the Space Age. Figures~\ref{fig7}--\ref{fig12} indicate that this conclusion hold not only when looking at solar activity as a whole but also when focusing specifically on solar wind conditions during SIR/HSSs. It is noteworthy that, while the IMF magnitude (see Figure~\ref{fig8}) had very low values during the late declining phase of SC23 as discussed above, the HSS peak velocity was meanwhile at its highest values, and started to drop only during the rising phase of SC24 (see Figure~\ref{fig7}) and remained at lower values than during the previous cycle. Even though the magnetic field has to some extent recovered from the deep minimum of SC23, it has continued to be at a lower level during SC24 than at the corresponding phases during SC23.

In fact, the early/late declining phase distinction for SC24 cannot presently be determined, as the minimum is still to be reached. Whether such a distinction will even be needed for SC24 is not certain, as it may be so that the unusually long declining phase of SC23 and the associated extremely weak activity conditions were unique.

\section{Conclusions}

We presented a method for detection of solar wind high-speed stream events, based on the IMF magnitude time derivative, the solar wind velocity, and, to remove ICMEs misidentified as SIR/HSSs and SIR/HSSs containing an embedded ICME, comparison with the Richardson and Cane list of ICMEs. This algorithm has been applied to years 1995--2017, leading to the detection of 588~SIR/HSS events with speeds exceeding 500~km/s. When compared to the existing HSS lists, which use slightly different criteria and may include a human visual detection of events, the algorithm proves efficient in correctly identifying SIR/HSS events and setting their beginning at the time when the solar wind speed starts increasing. The main benefit of this new detection method is that it produces a catalog of SIR/HSS events over 23~years of solar wind observations in a transparent and reproducible way.

The analysis of the 588~SIR/HSS events obtained with the detection method shows that the yearly number of SIR/HSS events peaks during the declining phase of the solar cycle. Besides the maximum yearly number of SIR/HSS events in SC23 (39~events in 2007) is about the same as the maximum number in 1995 of SC22 (35~events) and in 2016 of SC24 (35~events). It is yet to be seen if the yearly number of SIR/HSS events in SC24 increases when approaching the solar cycle minimum.

The number of SIR/HSS events sharply drops after the solar minimum is reached, during the first year of a new cycle. After that, the yearly number of events starts to rise. Even during the rising phases, which are associated with the smallest number of SIR/HSSs, the yearly number of SIR/HSSs amounts to 69\% and 48\% of the yearly number during the declining phases of SC23 and SC24, respectively. During the solar maximum years, which are dominated by ICME events, we find a local minimum in the occurrence of SIR/HSS events.

During the late declining phase of SC23, large low-latitude coronal holes persisted on the solar surface, which had long lifetimes. This resulted in repeated very-high-velocity streams. Even though the compressions of magnetic field were of the same magnitude as during the late declining phase of the previous SC22, the low background magnetic field values at 1~AU near the ecliptic plane during the extended minimum of 2007--2008 \citep[consistent with results by][]{Lee2009,Wu2013} resulted in low $B$ values in the ecliptic SIR regions and in low solar wind coupling function (Akasofu $\varepsilon$) values. Hence, the geoeffectiveness of SIR/HSSs during the late declining phase of SC23 was the smallest of this entire cycle.

During SC24, both the solar wind velocity and the magnetic field have continued to be at a lower level than during the corresponding phases of SC23. Hence, the geoeffectiveness of SIR/HSSs has been lower too. For auroral electrojet and substorm activity measured by the $AE$ index, the geoeffectiveness has been about 40\% lower during the rising phase and 20\% lower during the early declining phase. For the ring current and storms measured by the $SYM$--$H$ index, the geoeffectiveness has been about 35\% lower during the rising phase and 20\% lower during the early declining phase.

We can confirm the earlier-reported fact the SIR/HSS events have highest geoeffectiveness during the declining phases. However, there may be solar-cycle-dependent differences when one considers the early and late declining phases. The late declining phase of SC22 was very geoeffective, whereas the corresponding phase in SC23 was not. The early declining phase was the most geoeffective one inside SC23. So far, for SC24, the early declining phase has also been the most geoeffective one; however, we do not know yet how geoeffective SIR/HSSs will be during the late declining phase of SC24.

\acknowledgments
This work is supported by the Academy of Finland (projects 312351 and 285474) and the European Research Council (Consolidator Grant 682068-PRESTISSIMO).
The OMNI data were obtained from the GSFC/SPDF OMNIWeb interface at \url{http://omniweb.gsfc.nasa.gov}. The monthly values of the sunspot number between 1995 and 2017 were retrieved from the World Data Center SILSO, Royal Observatory of Belgium, Brussels. The Richardson and Cane list of ICMEs was downloaded from \url{http://www.srl.caltech.edu/ACE/ASC/DATA/level3/icmetable2.htm}. The solar cycle beginning dates were determined from \url{ftp://ftp.ngdc.noaa.gov/STP/space-weather/solar-data/solar-indices/sunspot-numbers/cycle-data/table_cycle-dates_maximum-minimum.txt}. This work utilizes the HSS catalog issued by G.~Mari{\cb s} Muntean, D.~Be{\cb s}liu--Ionescu and V.~Dobric{\u a}, managed by the Institute of Geodynamics of the Romanian Academy.

\listofchanges

\end{document}